\documentclass[longbibliography,noeprint,noshowpacs,nopreprintnumbers,twocolumn,pra,showpacs,superscriptaddress,amsmath,amssymb,citeautoscript,aps,10pt]{revtex4-2}
\usepackage{graphicx}
\usepackage{dcolumn}
\usepackage{color}
\usepackage{bm}
\usepackage{bbold}
\usepackage{soul}
\usepackage[hidelinks]{hyperref}
\hypersetup{
    colorlinks,
    citecolor=blue,
    filecolor=blue,
    linkcolor=blue,
    urlcolor=blue
}
\graphicspath{{Figures/}{./Figures/}{.}}

\DeclareMathOperator{\E}{\mathcal{E}}
\DeclareMathOperator{\B}{\mathcal{B}}
\DeclareMathOperator{\C}{\mathcal{C}}
\DeclareMathOperator{\hc}{\mathcal{H}}
\DeclareMathOperator{\hh}{\hat{H}}
\renewcommand{\vec}{\bm}

\begin{document}

\title{A Dynamical Bulk-Boundary Correspondence in Two Dimensional Topological Matter}

\author{Tomasz Mas{\l}owski}
\affiliation{The Faculty of Mathematics and Applied Physics, Rzesz\'ow University of Technology, al.~Powsta\'nc\'ow Warszawy 6, 35-959 Rzesz\'ow, Poland}
\author{Jesko Sirker}
\affiliation{Department of Physics and Astronomy and Manitoba Quantum Institute, University of Manitoba, Winnipeg, Canada R3T 2N2}
\author{Nicholas Sedlmayr}
\email[e-mail:]{sedlmayr@umcs.pl}
\affiliation{Institute of Physics, M. Curie-Sk\l{}odowska University, 20-031 Lublin, Poland}

\date{\today}

\begin{abstract}
We provide strong numerical evidence for a dynamical bulk-boundary correspondence in two-dimensional topological matter which manifests itself as boundary contributions to the dynamical free energy and is governed by a two-dimensional non-Hermitian dynamical Loschmidt matrix---a setting largely unexplored beyond one dimension. Following a quantum quench, in-gap bands emerge in the spectrum of the Loschmidt matrix between successive dynamical quantum phase transitions when the time-evolving Hamiltonian is topological, while they are absent for quenches into the trivial phase in all cases we have studied. By fitting these in-gap bands, we show that they account for the observed boundary contributions to the dynamical free energy thus supporting a direct connection between the spectrum of a non-Hermitian dynamical matrix and topological boundary contributions. Taken together with earlier studies of the one-dimensional case, our results provide a framework to understand and classify dynamical topological phenomena based on the spectral properties of certain non-Hermitian matrices.
\end{abstract}

\maketitle


\section{Introduction}\label{sec:intro}

Over the last decades an extension to the Landau picture of phase transitions has emerged which considers phases also defined by topology, rather than just by symmetries. Such topological phase transitions are quantum phase transitions accompanied by a change in a topological index~\cite{Ryu2010,Hasan2010,Trifunovic2021}. This index identifies which gapped band structures can be continuously deformed into each other under unitary transformations. Typically a topological phase transition is therefore accompanied by the closing of the bulk gap \cite{Maska2021}. If we restrict the unitary transformations to those which preserve certain non-local symmetries of the Hamiltonian then this is known as symmetry protected topological (SPT) order. Studying SPT order has gained widespread interest in large parts due to the bulk-boundary correspondence which closely relates the bulk topological index to gapless boundary modes~\cite{Chiu2016,Trifunovic2018}.

In a somewhat similar manner the Ehrenfest classification of phase transitions, whereby the order of the transition is which derivative of the free energy becomes discontinuous, has found an analogue in so-called dynamical quantum phase transitions (DQPTs)~\cite{Heyl2013,Andraschko2014,Heyl2018a,Heyl2019,Sedlmayr2019a}. In this case the dynamical free energy, also called the return rate, becomes non-analytic at critical \emph{times} during time evolution. The return rate is to the Loschmidt amplitude, for our purposes the overlap between a time evolved and the initial state, as the free energy to the partition function. Another curious correspondence occurs when considering the overlap between equilibrium states. Known as the fidelity it has been shown that this can be used to identify topological phase transitions and the topologically protected boundary modes arising due to the bulk-boundary correspondence~\cite{Sirker2014}. Previously it has been numerically demonstrated---based on an analysis of the return rate and the Loschmidt matrix---that a dynamical bulk-boundary correspondence (DBBC) exists for one dimensional systems and two dimensional higher order materials~\cite{Sedlmayr2018,Sedlmayr2019a,Maslowski2020,Maslowski2023,Maslowski2024}. However, these cases are special since the eigenvalues of the Loschmidt matrix responsible for the DBBC have no remaining degrees of freedom and are pinned to zero. Here we are taking a first step to establishing the connection between the spectrum of the non-Hermitian Loschmidt matrix and boundary contributions to the dynamical free energy in a more general setting by studying two-dimensional topological insulators and superconductors.

Non-equilibrium phenomena do not admit of the relatively easy classification of equilibria, and in order to cope with the wide landscape of possibilities particular forms of non-equilibrium behavior are often focused on. One such focus is on quenches in quantum models. In this case an initial state is prepared, often the ground state of a Hamiltonian $\hat H_0$, and then it is time evolved ``suddenly'' with a different Hamiltonian $\hat H_1$, normally related to the initial Hamiltonian by a change in some parameters. The resulting dynamics can be studied in various ways and has been focused on for example in the field of quantum thermalization~\cite{Rigol2008a,Polkovnikov2011,Rigol2012,Sirker2014}. Here we focus exclusively on such quench scenarios. Dynamical quantum phase transitions themselves can be defined in several ways. First one can consider the dynamics of an order parameter as a function of some external parameter and when it becomes zero~\cite{Marino2022,Corps2022}, or one can look at non-analyticities of the return rate during dynamics~\cite{Heyl2013,Heyl2018a}. Though some connections between these exist, it is not currently clear if they are always related, and on the face of it they seem distinct. Here we focus entirely on the second case, sometimes referred to as DQPT type II. These DQPTs allow for a reasonably systematic review of certain non-equilibrium phenomena which nonetheless contain physics not captured by the equilibrium phase diagrams for the initial or time evolving Hamiltonians~\cite{Vajna2014,Andraschko2014,Karrasch2017,Jafari2017,Cheraghi2018,Jafari2019,Wrzesniewski2022}. We would also stress that although we have a critical point, now occurring at a time, there does not always appear to be any clear sense in which this necessarily divides different \emph{dynamical} phases.

A large amount of work already exists on DQPTs both theoretically~\cite{Karrasch2013,Heyl2014,Heyl2015,Sharma2015,Halimeh2017,Homrighausen2017,Halimeh2018,Shpielberg2018,Zunkovic2018,Srivastav2019,Huang2019,Gurarie2019,Abdi2019,Puebla2020,Link2020,Sun2020,Rylands2021,Trapin2021,Yu2021,Halimeh2021,Halimeh2021a,DeNicola2021,Cheraghi2021,Cao2021,Bandyopadhyay2021,Cheraghi2023,Wong2023b} and experimentally~\cite{Jurcevic2017,Flaschner2018,Zhang2017b,Guo2019,Smale2019,Nie2020,Tian2020}. Additionally, DQPTs in Floquet systems have been considered~\cite{Sharma2014,Yang2019,Zamani2020,Zhou2021a,Jafari2021,Hamazaki2021,Zamani2022,Luan2022,Jafari2022} and the idea has been generalized to mixed states, finite temperatures, and dissipative systems~\cite{Mera2017,Sedlmayr2018b,Bhattacharya2017a,Heyl2017,Abeling2016,Lang2018,Lang2018a,Kyaw2020,Starchl2022,Naji2022,Kawabata2023,Jafari2025}. Of particular interest for us here is the focus on DQPTs in topological matter~\cite{Schmitt2015,Vajna2015,Jafari2016,Bhattacharya2017,Jafari2017a,Sedlmayr2018,Jafari2018,Zache2019,Maslowski2020,Okugawa2021,Rossi2022,Maslowski2023,Maslowski2024} where attempts to define dynamical order parameters have also been made~\cite{Budich2016,Heyl2017,Bhattacharya2017a}. Work extending beyond topological models in one dimension~\cite{Schmitt2015,Vajna2015,Bhattacharya2017,DeNicola2022,Hashizume2022,Brange2022,Sacramento2024,Kosior2024,Maslowski2024b} or simple two band models~\cite{Huang2016,Jafari2019,Mendl2019,Maslowski2020,Maslowski2023,Maslowski2024,Maslowski2024b} remains relatively under-explored.

By extending time to the complex plane one can investigate the zeroes of the Loschmidt echo, i.e.~the Fisher zeroes. DQPTs are related to when they cross or touch the real time axis~\cite{Maslowski2024b}. In one dimension the Fisher zeroes form lines leading to critical times at which the return rate is non-analytic \cite{Heyl2013}. In two dimensions, which we now focus on here, the lines generically become areas of Fisher zeroes with a varying density. For simple two band models one can then easily show that critical times occur at the edges of the Fisher zero areas as they cross the real time axis~\cite{Schmitt2015,Vajna2015,Maslowski2024b}, leading to cusps in the first derivative of the return rate. More generically it can be shown that cusps always occur at the edges of the areas of Fisher zeroes, and potentially inside, with details of the cusp depending on the density of the zeroes. This analysis applies equally to higher dimensions~\cite{Maslowski2024b}.

The DBBC is the correspondence between significant periodic contributions to the boundary return rate and the nature of the quench between topological phases~\cite{Sedlmayr2018}. Such contributions are observed when the time evolving Hamiltonian is in the topological phase, whereas they are absent for a time evolving Hamiltonian which belongs to the trivial phase. In the one-dimensional case, it has been shown that these boundary contributions stem from exponentially small eigenvalues of the non-Hermitian dynamical Loschmidt matrix \cite{Sedlmayr2018}. These exponentially small eigenvalues periodically appear in its otherwise gapped bulk spectrum. However a full theory relating this phenomenon to topological invariants has not been worked out yet. We note that the dynamical topological indices previously defined~\cite{Budich2016,Heyl2017,Bhattacharya2017a} do not predict the occurence of the exponentially small eigenvalues~\cite{Sedlmayr2018}. While it has been shown that the same connection between the spectrum of the Loschmidt matrix and the boundary contributions to the dynamical free energy also holds for higher order topological insulators (HOTIs) in two dimensions~\cite{Maslowski2023,Maslowski2024} this case is still similar to the one-dimensional one in the sense that the eigenvalues have no dispersion and are pinned to zero.

An important question is therefore whether the connection between the spectrum of the dynamical Loschmidt matrix and boundary contributions to the free energy holds more generally. A natural first step is to investigate two-dimensional topological matter where eigenvalues have a momentum degree of freedom and are thus no longer expected to be pinned to zero. Instead, one might expect---if the connection holds more generally---that there are now bands of in-gap modes which show up periodically in the Loschmidt spectrum. As one of our main results we find that this is indeed the case. Furthermore we are able to relate the boundary return rate directly to the in-gap modes and show how one is caused by the other. Also, in contrast to the one-dimensional case, the dynamically changing in-gap band results in a boundary return rate which is no longer a simple plateau between successive critical times. We emphasize that the reported results occur consistently over a variety of models we investigated and, as we will demonstrate, are stable against the breaking of translational invariance. However, a full characterization of this phenomenon in terms of a non-Hermitian topological invariant is currently lacking and remains an open problem.

This paper is organized as follows. In section \ref{sec:dqpt} we introduce DQPTs and the dynamical bulk-boundary correspondence. In section \ref{sec:mods} we define the exemplary model we will study. In section \ref{sec:kitdbb} we show the results for the dynamical bulk-boundary effect, and discuss its general properties. In section \ref{sec:con} we conclude.

\section{Dynamical Quantum Phase Transitions}\label{sec:dqpt}

A dynamical quantum phase transition occurs when the return rate becomes non-analytic as a function of time. We start by defining an initial pure state $|\Psi_0\rangle$ which is typically, and will be here for us, the ground state of a Hamiltonian: $\hh_0|\Psi_0\rangle=E_0|\Psi_0\rangle$. This state is then time evolved by a second Hamiltonian $\hh_1$ and the overlap between the time-evolved and original state, the Loschmidt amplitude, is defined as
\begin{equation} \label{LN}
L_{N_x\times N_y}(t)=\langle\Psi_0|e^{-i\hh_1t}|\Psi_0\rangle\,,
\end{equation}
for a system of size $N_x\times N_y$. The Loschmidt \emph{echo} in turn is then defined as $|L_{N_x\times N_y}(t)|^2$. The Loschmidt echo suffers an orthogonality catastrophe in the thermodynamic limit motivating the definition of the return rate
\begin{equation}\label{return}
l_{N_x\times N_y}(t)=-\frac{1}{N_x N_y}\ln|L_{N_x\times N_y}(t)|
\end{equation}
which plays the role of a dynamical free energy. We define $l(t)\equiv\lim_{N_x,N_y\to\infty}l_{N_x\times N_y}(t)$ in the thermodynamic limit. The return rate $l(t)$ can become non-analytic at critical times, which are called DQPTs. The nature of the non-analyticity depends on the density of zeroes of $L(t)$ which occur at particular times or lengths of time. For thermodynamic phase transitions, an understanding of why non-analyticities in thermodynamic quantities occur can be obtained by considering the zeroes of the partition function either as a function of a complex chemical potential or as a function of a complex inverse temperature. The former are known as Lee-Yang zeroes while the latter are called Fisher zeroes \cite{Bena2005,Lee1952,Yang1952,Fisher1965}. In both cases their distribution in the complex plane and how they approach the real axis determines the properties of the phase transition. In a similar manner, one can consider $L(-iz)$ for complex $z$, its zeroes are also typically called Fisher zeroes and form curves or surfaces in the complex plane. In one dimension the Fisher zeroes are lines and lead to a cusp in the return rate when they cross or touch the real time axis. In two dimensions the Fisher zeroes are generically areas typically leading to cusps in the return rate derivative, $\dot l(t)=\partial l/\partial t$, which occur at the edges of the areas of Fisher zeroes where they cross the real time axis \cite{Schmitt2015,Vajna2015,Maslowski2024b}.

The Loschmidt amplitude can be calculated from the correlation matrix $\C_{ij}=\langle\Psi_0|\Psi^\dagger_i\Psi_j|\Psi_0\rangle$, using~\cite{Levitov1996,Klich2003,Rossini2007}
\begin{equation}\label{rle}
    L(t)=\det{\bm M}(t)\equiv\det\left[1-\mathbf{\C}+\mathbf{\C}e^{-it {\mathcal H}^1}\right]\,.
\end{equation}
Here $i$ and $j$ simply label some set of basis states of the Hilbert space of $\hh_{0,1}$, with $\Psi^{(\dagger)}_j$ their annihilation (creation) operators, and $[\mathcal{H}^{0,1}]_{ij}=\langle i|\hh_{0,1}|j\rangle$ is the Hamiltonian density. We refer to $\bm M(t)$ as the Loschmidt matrix, and some of its properties are given in appendix \ref{app:prop}. The non-Hermitian Loschmidt matrix has eigenvalues $\lambda_i(t)$, in terms of which one finds
\begin{equation}\label{ele}
    L(t)=\prod_i\lambda_i(t).
\end{equation}
We will use the convention throughout that $\lambda_0(t)$ refers to whichever eigenvalue has the lowest magnitude at time $t$, and similarly $\lambda_{0,k}(t)$ for the lowest magnitude eigenvalue at some momentum $k$.

If we have a two band model which is diagonal in reciprocal space $\vec k$ then considerable simplifications can be made. Let the single particle eigenenergies of $\hh_1$ be $\pm\epsilon^1_{\vec{k}}$ and the Hamiltonian density be parameterized as $\hc^{0,1}=\vec{d}^{0,1}_{\vec k}\cdot\vec{\sigma}$ with $\sigma^\alpha$ ($\alpha=x,y,z$) the Pauli matrices. Then the momentum resolved correlation matrix is
\begin{equation}
    \mathcal{C}_{\vec{k}}=\frac{1}{2}\left(1-\hat{\vec{d}}^0_{\vec{k}}\cdot\vec{\sigma}\right)
\end{equation}
with $\hat{\vec d}^{0,1}_{\vec{k}}$ unit normalized vectors and
\begin{equation}
    e^{-it\mathcal{H}^1_{\vec{k}}}=
    \cos(\epsilon^1_{\vec{k}}t)+i\hat{\vec{d}}^1_{\vec{k}}\cdot
    \vec{\sigma}\sin(\epsilon^1_{\vec{k}}t).
\end{equation}
The eigenvalues of the Loschmidt matrix are
\begin{equation}
    \lambda_i\to(1,\lambda_{\vec{k}})
\end{equation}
with
\begin{equation}
\lambda_{\vec{k}}=\cos(\epsilon^1_{\vec{k}}t)+i\hat{\vec{d}}^0_{\vec{k}}\cdot\hat{\vec{d}}^1_{\vec{k}}\sin(\epsilon^1_{\vec{k}}t)\,.
\end{equation}
That half the eigenvalues are equal to one, in the half-filled case we will focus on, follows from the structure of the correlation matrix, see appendix \ref{app:prop}. This leads to the known result~\cite{Vajna2015}
\begin{equation}\label{pbcloschmidt}
L_N(t)=\prod_{\vec{k}}\lambda_{\vec{k}}(t)=\prod_{\vec{k}}\left[\cos(\epsilon^1_{\vec{k}}t)+i\hat{\vec{d}}^0_{\vec{k}}\cdot\hat{\vec{d}}^1_{\vec{k}}\sin(\epsilon^1_{\vec{k}}t)\right]\,.
\end{equation}

For two-band models in any dimension the Fisher zeroes are thus simply~\cite{Vajna2015}
\begin{equation}
z_n({\vec{k}})=i\frac{\pi}{\epsilon^1_{\vec{k}}}\left(n+\frac{1}{2}\right)-\frac{1}{\epsilon^1_{\vec{k}}}\tanh^{-1}\left(\hat{\vec{d}}^0_{\vec{k}}\cdot\hat{\vec{d}}^1_{\vec{k}}\right) \label{zn}
\end{equation}
where $n$ is an integer. For a two dimensional model we have areas of zeroes in the complex plane, which give rise to critical times along the real time axis when the density of the zeroes either diverges or has a step like behavior~\cite{Maslowski2024b}. This results in non-analytic behavior for the return rate at the critical times
\begin{equation}\label{critt}
t^{1,2}_{c,n}=(\min,\max)\left[\frac{\pi}{\epsilon^1_{{\vec{k}}^*}}\left(n+\frac{1}{2}\right)\right]
\end{equation}
where the ${\vec{k}}^*$ satisfy the condition
\begin{equation}\label{condition}
\vec{d}^0_{{\vec{k}}^*}\cdot\vec{d}^1_{{\vec{k}}^*}=0\,.
\end{equation}
There can be additional critical times in between depending on details of the quenches~\cite{Maslowski2024b}. Some generalizations are also known to more complicated models~\cite{Maslowski2020,Maslowski2023,Maslowski2024}, but generically once we no longer have a two band model, and we cannot take advantage of reciprocal space, we must use Eq.~\eqref{rle} to obtain the Loschmidt amplitude and rely on numerical calculations.

The DBBC manifests itself in the boundary contributions to the return rate. In two dimensions we can compare two different geometries: open boundary conditions along one direction (``ribbons''), or along both (``flakes''). For the two dimensional Kitaev lattice model, which we will focus on - see next section, this can be an instructive comparison as the ribbon geometry allows the existence of exponentially small energy states along each side which become gapped for the fully open flake system. Perhaps surprisingly these zero modes have a clear signature in the boundary return rate, even compared to the in-gap bands. For the ribbon geometry with periodic boundary conditions in x-direction and open boundary conditions in y-direction the return rate will scale as 
\begin{equation}\label{bbreturn}
    l_{N_x\times N_y}(t)\sim l(t)+\frac{2l_{B}(t)}{N_y}+\frac{A(t)}{N_x N_y}\,.
\end{equation}
For the flake we only consider the case $N=N_x=N_y$ in which case the return rate scales as 
\begin{equation}\label{bbreturno}
    l_{N_x\times N_y}(t)\sim l(t)+\frac{4l_{B}(t)}{N}+\frac{A(t)}{N^2}\,,
\end{equation}
with $l(t)$ and $l_{B}(t)$ the bulk and boundary contributions respectively and $A(t)$ the lowest order bulk correction. The factor of 2 between the definitions of the boundary contributions for the two geometries occurs because the ribbon has two spatially separated edges while a square flake has 4 connected edges. The bulk contribution can be accurately found from the reciprocal space form, where larger system sizes are possible, and we find the boundary term from a finite size scaling analysis. 

It has been shown that in one dimension the dynamical bulk-boundary effect is caused by the existence of eigenvalues of the Loschmidt matrix which become pinned to zero between successive critical times~\cite{Sedlmayr2018,Maslowski2020}. In turn these zero modes cause large plateaus in the boundary return rate $l_B(t)$. This still holds for higher order topological systems in two dimensions, in which the edge modes are still localized zero energy modes, although in this case the critical times are once again typically extended regions rather than points as the bulk system is two dimensional~\cite{Maslowski2023,Maslowski2024}. However, there can be exceptions to this in which the critical times are still strictly points and Fisher zeroes form lines~\cite{Maslowski2024}. These results suggest that the boundary contribution might serve as an order parameter separating different dynamical phases with a DQPT between these phases. However, it is not a priori clear if these results also hold in more general cases where the eigenvalues of the Loschmidt matrix acquire a dispersion. Answering this question is crucial to understand whether the connection between the spectrum of the dynamical non-Hermitian Loschmidt matrix and boundary contributions is a general principle or rather a consequence of the simple topology of the systems considered so far which pins eigenvalues to zero.

We note that in general the Chern number and bulk-boundary correspondence~\cite{Teo2010} tells one only the minimum number of bands crossing the gap, but not whether there are Majorana zero modes present or indeed exactly where and how the bands cross the gap~\cite{Sedlmayr2017,Glodzik2023}. For a system on a cylindrical geometry, \emph{i.e.}~when we have open boundaries along one direction and periodic along the perpendicular direction, topological bands which cross at time reversal invariant momenta have Majorana zero modes at zero energy. The number of these modes can in general depend on the orientation of the edges~\cite{Sedlmayr2017,Glodzik2023}, and they naturally do not occur at all if there is only one single edge. As we will demonstrate we find a natural analogue of this phenomena in the Loschmidt spectra and boundary return rate. In particular, we show that the extracted boundary return rate can be directly attributed to the in-gap modes.

\section{An Exemplary Model}\label{sec:mods}

\begin{figure}[t!]
\includegraphics[height=0.85\columnwidth]{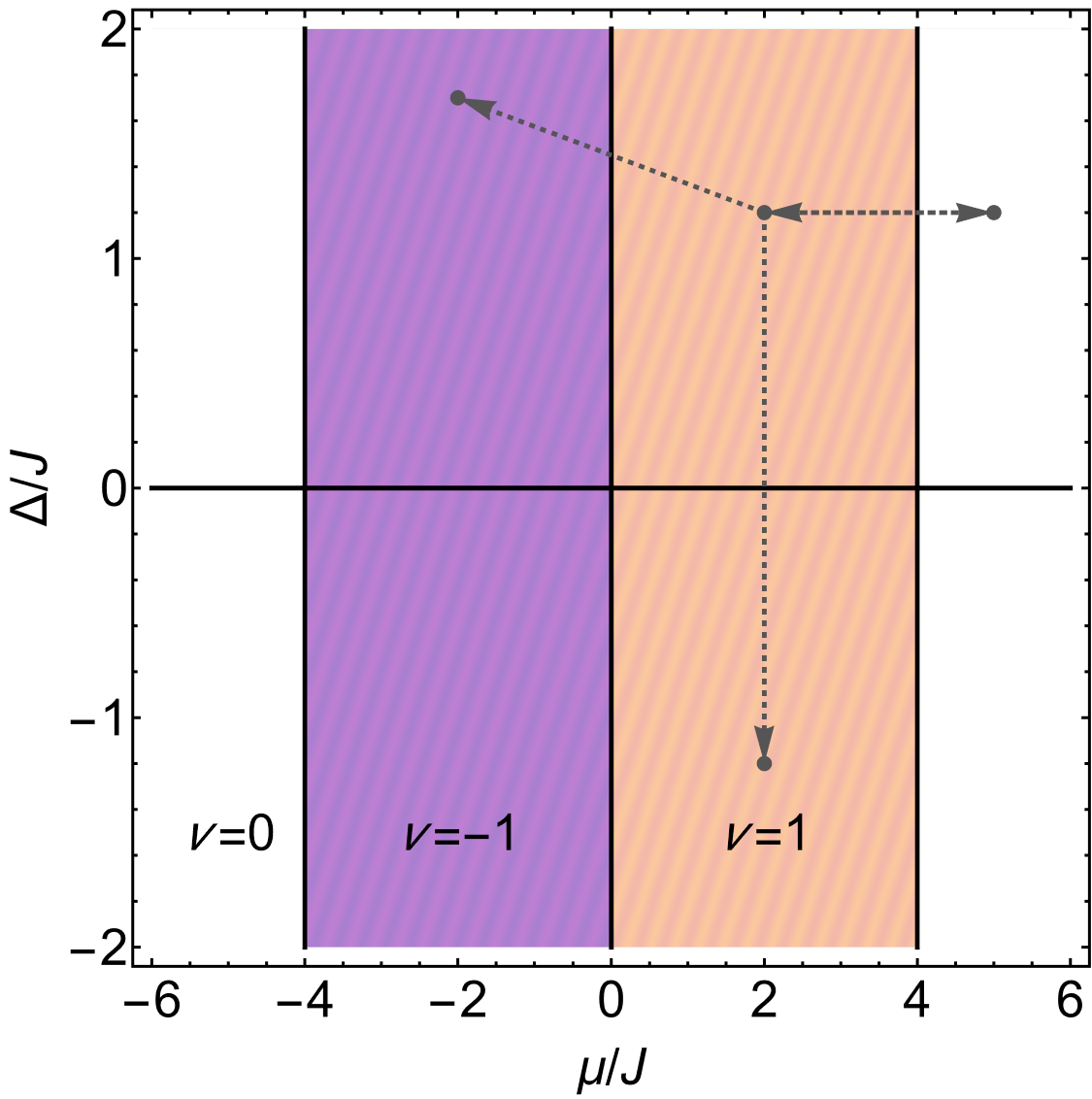}
\caption{The topological phase diagram of the two dimensional Kitaev model, see Eq.~\eqref{kit_ham}. Bulk gap closings are shown as solid lines. Dashed lines with arrows show the quenches considered in this article, see the main text for parameters.} 
\label{fig:phase}
\end{figure}

Throughout this paper we will illustrate our findings using an exemplary topological superconductor, though all results extend naturally and directly to topological insulators. We use a paradigmatic two band model, for which many analytical results may be obtained, and which demonstrates clearly the basic results. This is a two dimensional generalization of the Kitaev chain~\cite{Kitaev2001,Sedlmayr2015b}, which becomes a spinless $p_x+ip_y$ superconductor on a square lattice. This model has a very simple equilibrium topological phase diagram with available Chern numbers $\nu\in\{-1,0,1\}$ and only two independent parameters. However we note that for more complicated models where quenches between a wider range of Chern numbers are possible, or with different kinds of boundary, it is not clear if all results directly generalize. This is the subject of ongoing investigations. In general we have no reason to expect a direct mapping between the behavior of the equilibrium phase diagrams and the dynamics, particularly when quenching between higher Chern numbers.

The two dimensional Kitaev lattice is given by
\begin{align}\label{kit_ham}
\hat{H}=&
\sum_{j,\ell}\Psi^\dagger_{j,\ell}\left(
\Delta i\sigma^y-J\sigma^z\right)\Psi_{j+1,\ell}+\textrm{H.c.}
\nonumber\\\nonumber&
+\sum_{j,\ell}\Psi^\dagger_{j,\ell}\left(
\Delta i\sigma^x-J\sigma^z\right)\Psi_{j,\ell+1}+\textrm{H.c.}
\\
&-\mu\sum_{j,\ell}\Psi^\dagger_{j,\ell}\sigma^z\Psi_{j,\ell}\,.
\end{align}
where $\Psi^\dagger_{j,\ell}=\{c^\dagger_{j,\ell},c_{j,\ell}\}$ with $c_{ j,\ell}^{(\dagger)}$ annihilating (creating) a spinless particle at site $(j,\ell)$ on a square lattice. $J$ is the hopping strength and $\Delta$ the p-wave pairing strength with $\mu$ the chemical potential. $\sigma^\alpha$ are the Pauli matrices representing particle-hole space. Following a Fourier transform the Hamiltonian becomes
\begin{align}\label{kit_hamk}
\hat{H}=&\sum_{{\vec{k}}}\Psi^\dagger_{{\vec{k}}}\left(-2J\cos k_x-2J\cos  k_y-\mu\right)\sigma^z\Psi_{{\vec{k}}}
\\\nonumber&
-2\Delta\sum_{{\vec{k}}}\Psi^\dagger_{{\vec{k}}}\left(\sin k_x\sigma^y+\sin k_y\sigma^x\right)\Psi_{{\vec{k}}}
\\\nonumber\equiv&
\sum_{{\vec{k}}}\Psi^\dagger_{{\vec{k}}}{\bm d}_{{\vec{k}}}\cdot\vec{{\bm \sigma}}\Psi_{{\vec{k}}}\equiv
\sum_{{\vec{k}}}\Psi^\dagger_{{\vec{k}}}\hc_{{\vec{k}}}\Psi_{{\vec{k}}}\,
\end{align}
which has eigenvalues $\pm\epsilon_{{\vec{k}}}$ given by
\begin{align}
	\epsilon^2_{{\vec{k}}}=&\left(2J\cos k_x+2J\cos k_y+\mu\right)^2\\\nonumber&+4\Delta^2\left(\sin^2k_x+\sin^2k_y\right).
\end{align}
This model possesses only a particle-hole symmetry $\mathcal{P}$,
\begin{equation}
	\mathcal{P}\hc_{{\vec{k}}}\mathcal{P}=-\hc^*_{-{\vec{k}}}\, ,
\end{equation}
where $\mathcal{P}^2=1$. For the Kitaev lattice defined here $P=\sigma^y$. Therefore it is in the D class of the topological periodic table with a $\mathbb{Z}$ invariant~\cite{Ryu2010,Chiu2016}, the Chern number~\cite{Thouless1982}. The phase diagram is straightforward to calculate using the Chern number, see Fig.~\ref{fig:phase}, on which we also mark the quenches we consider. The bandstructures of some of the topological points can be found in appendix \ref{app:bands}. The parameters used for the various exemplary quenches considered are: $\nu=0$ with $(\mu/J,\Delta/J)=(5,1.2)$, $\nu=1$ with $(\mu/J,\Delta/J)=(2,1.2)$, $\nu=-1$ with $(\mu/J,\Delta/J)=(-2,1.7)$, and $\nu=1$ with $(\mu/J,\Delta/J)=(2,-1.2)$. When this last point in parameter space is considered we will make explicit that it is the case with negative pairing $\Delta$. We note also that the phase of $\Delta$ can be changed continuously without crossing a topological phase boundary.

\begin{figure}
\includegraphics[width=0.95\columnwidth]{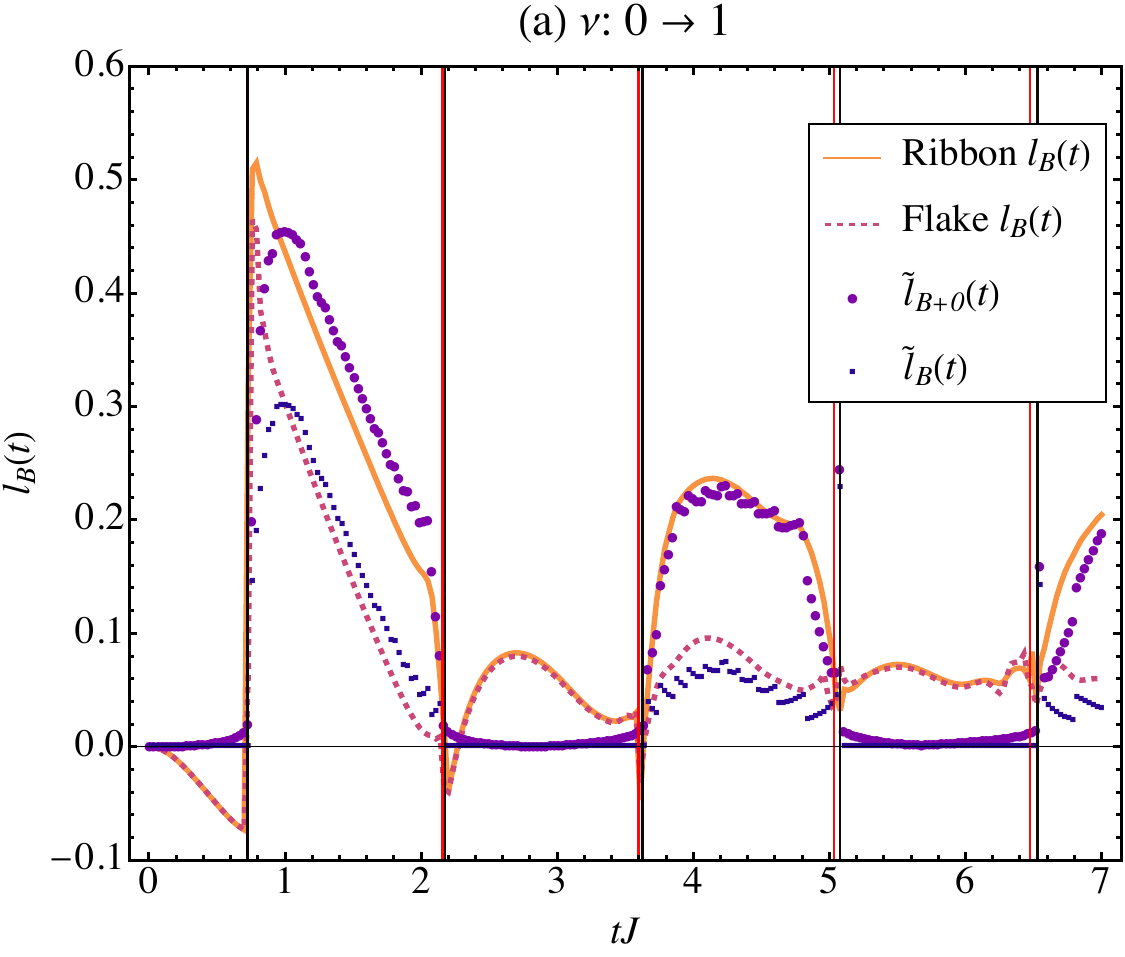}\\
\includegraphics[width=0.95\columnwidth]{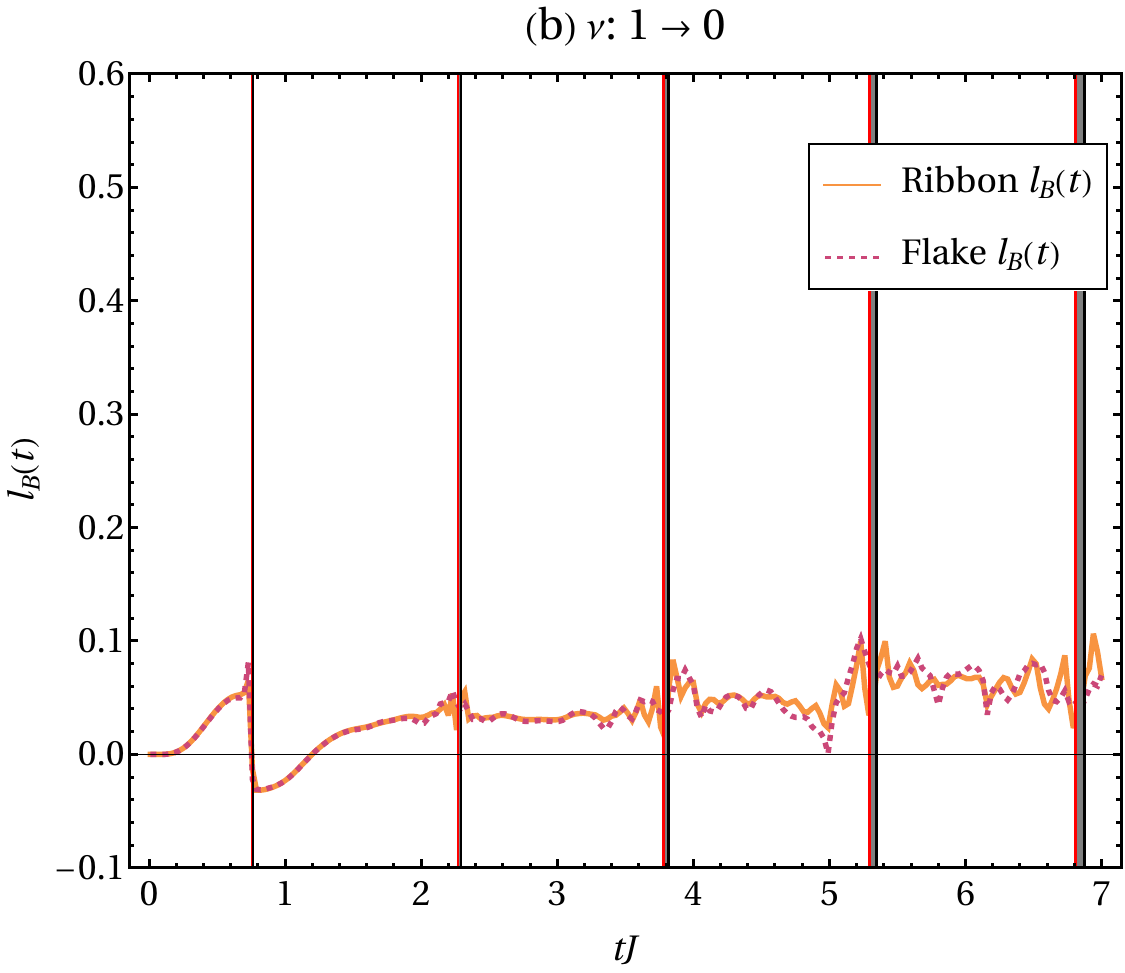}
\caption{Boundary return rate $l_B(t)$ for the Kitaev lattice extracted from a finite size scaling analysis. Vertical red and black lines show the beginnings and ends of critical regions respectively. The scaling, see Eqs.~\eqref{bbreturn} and \eqref{bbreturno}, is performed for ribbons of size $N_x\times N_y$ where $N_x=202$ is the length along the periodic direction, and $N_y\in\{200,300,\ldots,700\}$. For the flakes we use linear dimensions $N\in\{20,30,\ldots,80\}$. 
Also plotted in panel (a) are the contributions $\tilde l_{B}$ and $\tilde l_{B+0}$ due to the in-gap bands only, see Eqs.~\eqref{bound_an} and \eqref{bound_an0} respectively. Here the difference between the two geometries is an exponentially small Loschmidt eigenvalue present only for the ribbon.} 
\label{fig:br}
\end{figure}

\begin{figure*}
\includegraphics[width=0.95\columnwidth]{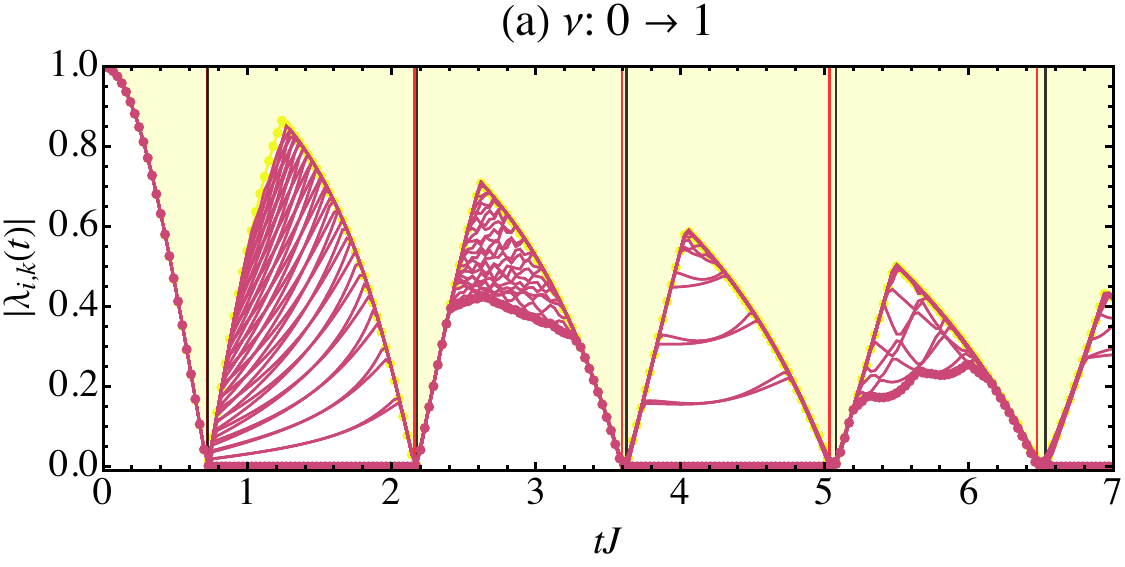}
\includegraphics[width=0.95\columnwidth]{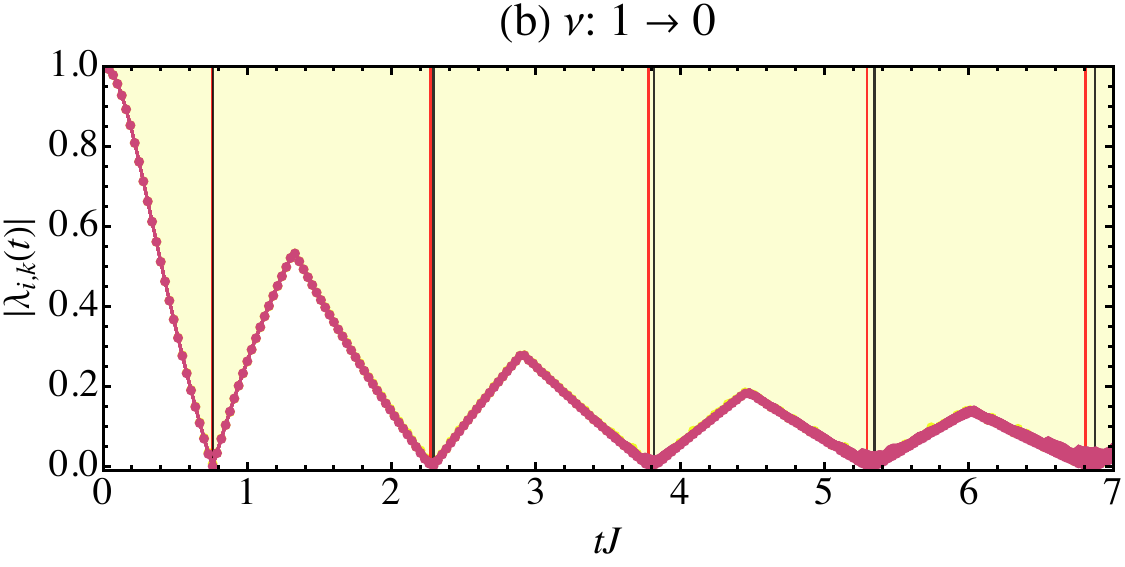}\\
\includegraphics[width=0.95\columnwidth]{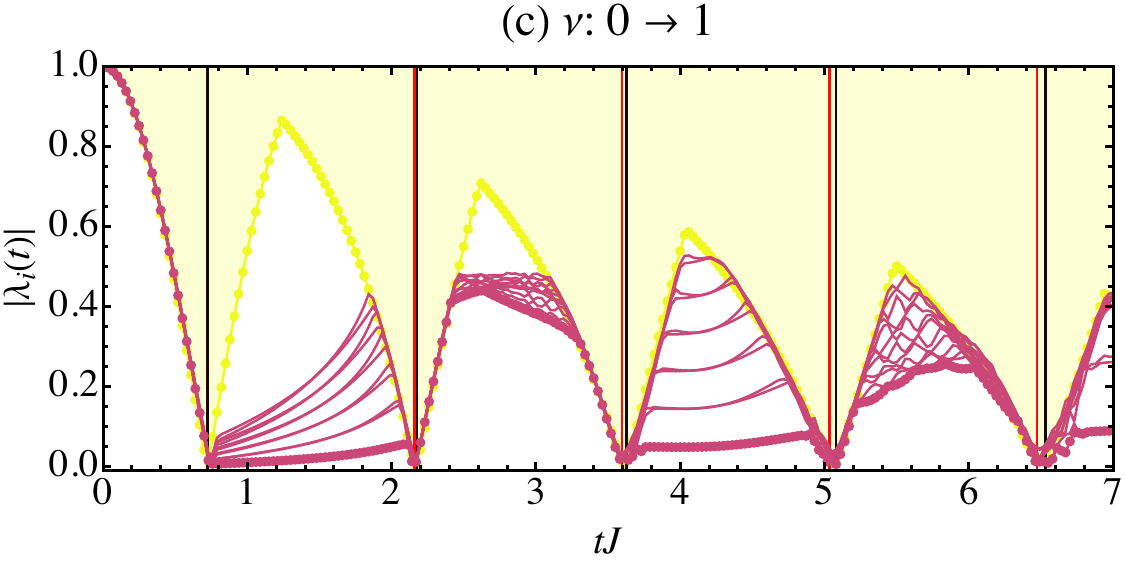}
\includegraphics[width=0.95\columnwidth]{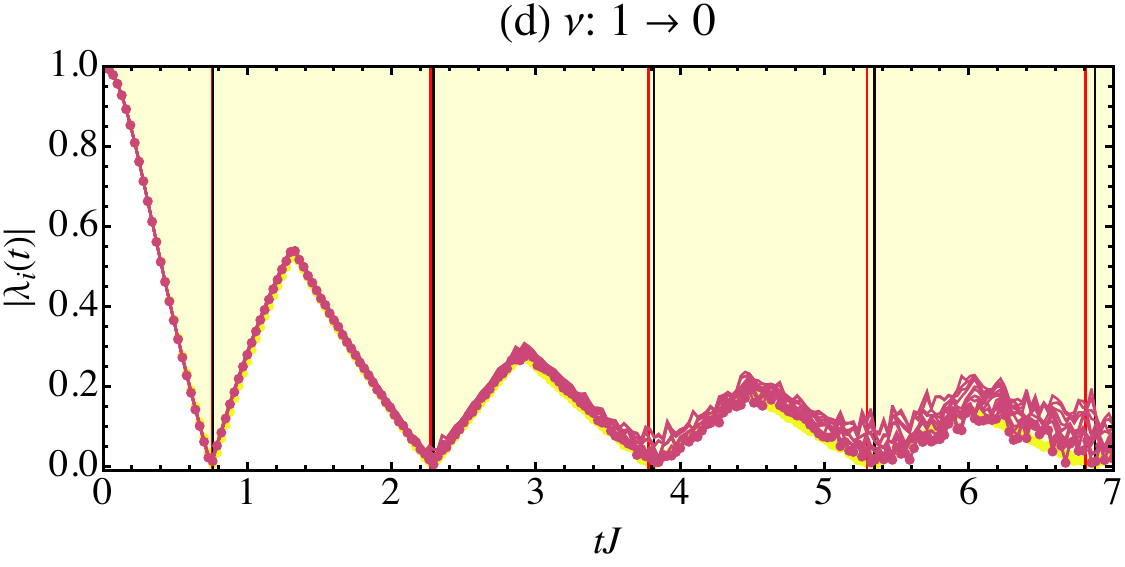}\\
\caption{(a,b) The 80 Loschmidt eigenvalues $|\lambda_i(t)|$  of smallest magnitude for a Kitaev ribbon of size $202\times 700$. (c,d) 12 smallest Loschmidt eigenvalues $|\lambda_i(t)|$ for a Kitaev flake of size $N=80$. The shaded regions show the areas where there are bulk eigenvalues. Vertical red and black lines show the beginnings and ends of critical regions respectively. The quenches $\nu\!:0\to1$ show the in-gap eigenvalues forming between successive critical times for both geometries. Zero eigenvalues are present for the ribbon but not the flake. For the quenches $\nu\!:1\to0$ no in-gap modes are present, in agreement with the DBBC, see also Fig.~\ref{fig:br}.} 
\label{fig:evr}
\end{figure*}

\section{Dynamical bulk-boundary Correspondence}
\label{sec:kitdbb}

The DQPTs for the Kitaev model \eqref{kit_ham} have been well studied in the bulk~\cite{Maslowski2024b}, and here we will simply summarize the main results of interest. Regions of Fisher zeroes cross the real time axis for quenches between all phases, resulting in critical times. It is also possible to cause DQPTs by quenching in the sign (and magnitude) of $\Delta$, which quenches between homeomorphic phases. Such a quench demonstrates that in general the analysis of DBBC cannot be linked solely to equilibrium topological phases. Analytical expressions for the critical times can also be found~\cite{Maslowski2024b}. We pick four possible quenches to demonstrate results for DQPTs between the various phases.

Of paramount interest is the comparison between the quenches $\nu\!:0\to1$ and $\nu\!:1\to0$. Based on previous results~\cite{Sedlmayr2018,Maslowski2020,Maslowski2023} we expect a significant dynamical boundary term only for the first of these. Indeed this is what we see, see Fig.~\ref{fig:br}, where a sizable boundary contribution exists between successive critical regions for the quench into the topologically non-trivial phase only. The difference between the two quenches is particularly obvious in the time interval $0.8\lesssim t \lesssim 2.2$ where the boundary contribution for the $\nu\!:0\to1$ quench is more than one order of magnitude larger than for the quench in the opposite direction. The large boundary contributions for the former quench can be seen for both the flake and the ribbon, however one distinction between these does remain. When there is a dynamical boundary term present in the ribbon, \emph{i.e.}~between successive critical regions when quenching \emph{into} the topologically non-trivial phase, $l_B(t)$ is larger than in the flake. We will return to this point below.

\subsection{Origin of the DBBC}
\label{Sec_Majorana}
For now, we would like to argue that the origin of the dynamical boundary term are in-gap eigenvalues of the Loschmidt matrix. In Fig.~\ref{fig:evr} the Loschmidt spectra of the ribbons and flakes are shown for the quenches $\nu\!:0\leftrightarrow1$. For the $\nu\!:0\to1$ quench, additional bands of Loschmidt eigenvalues inside the bulk gapped region are clearly visible, see Fig.~\ref{fig:evr}(a,c). By fitting these in-gap bands, see Eqs.~\eqref{bound_an} and \eqref{bound_an0}, we are able to show that they quantitatively account for the boundary contribution $l_B(t)$ (see symbols in Fig.~\ref{fig:br}). The quench into the trivial phase on the other hand shows no in-gap eigenvalues, see Fig.~\ref{fig:evr}(b,d). Further quenches between topologically non-trivial phases can be found in appendix \ref{app:dbb} and the results are consistent with the example discussed above. We have also checked that these in-gap eigenvalues are robust to disorder, see appendix \ref{app:disorder}, further supporting a topological origin.

While for the ribbon additional exponentially small eigenvalues appear periodically no such modes exist for the flake. We conjecture that this is due to the lack of zero energy modes in the spectra of the time evolving Hamiltonian, which are gapped out for the flake unlike for the ribbon due to the Majorana zero modes mixing ~\cite{Sedlmayr2017,Glodzik2023}. As we will show in more detail below, this difference is responsible for the different size of $l_B(t)$ for the two geometries as shown in Fig.~\ref{fig:br}.

\begin{figure}
\includegraphics[height=0.405\columnwidth]{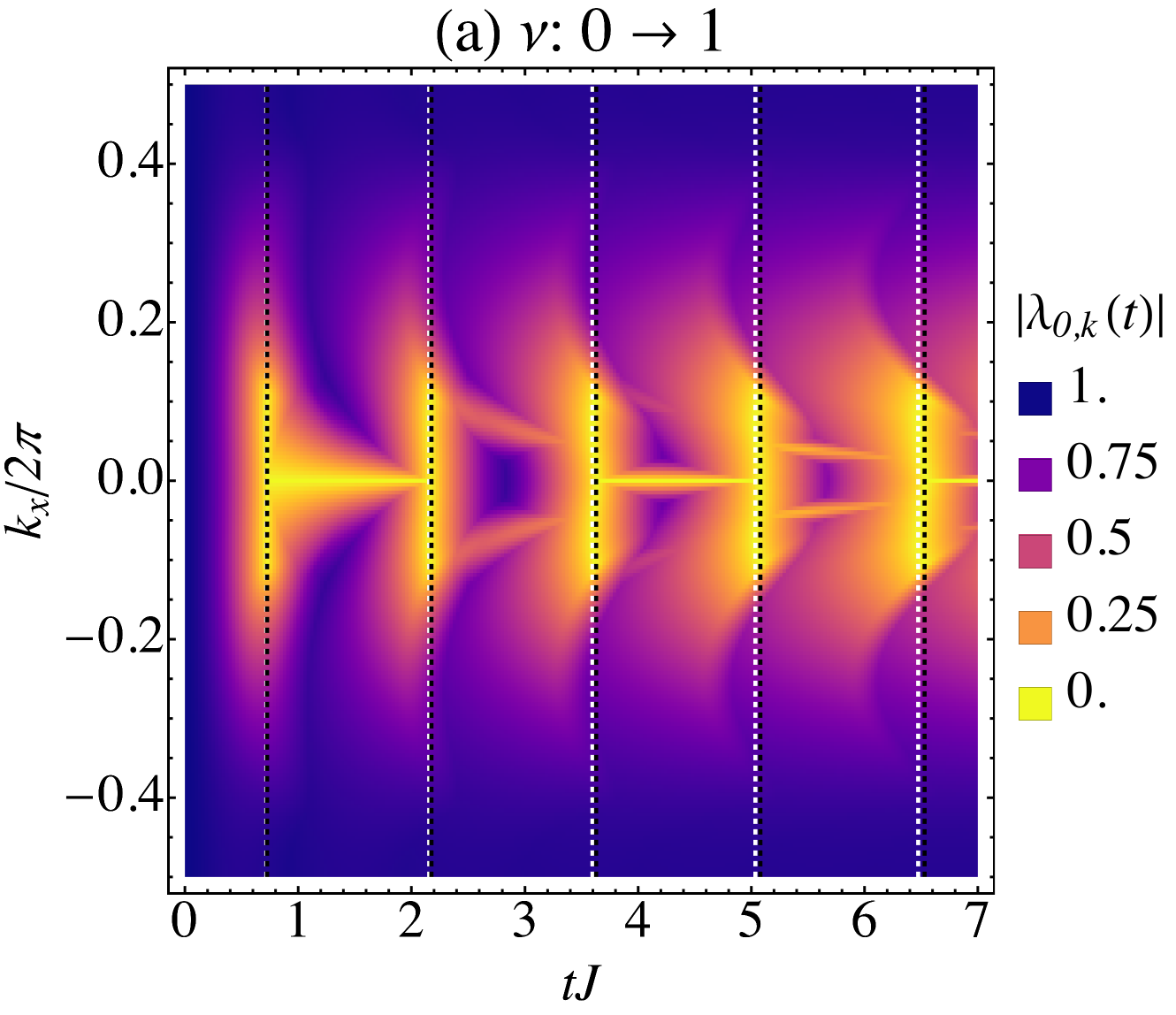}
\includegraphics[height=0.405\columnwidth]{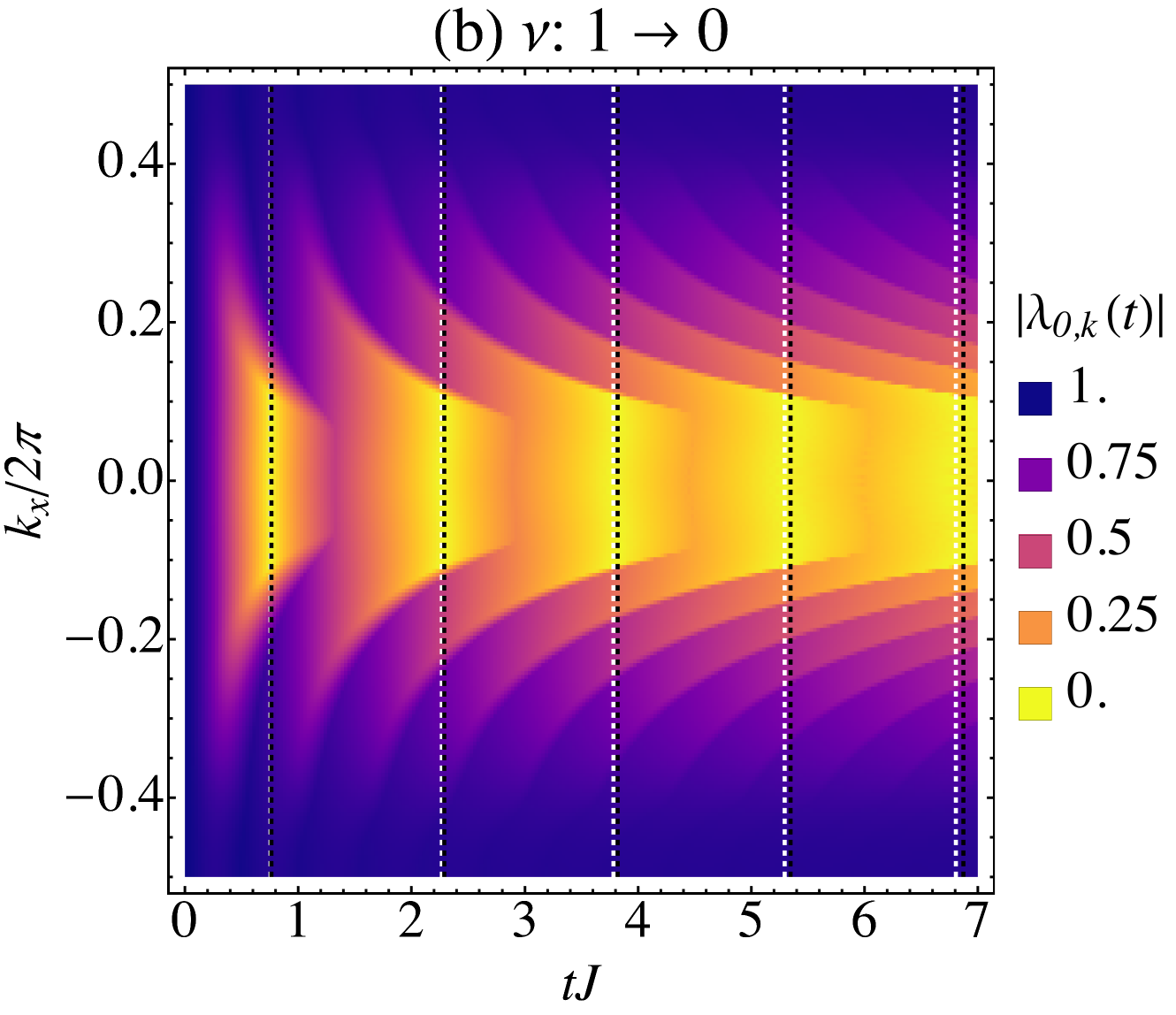}
\caption{The smallest Loschmidt eigenvalue $|\lambda_0(t)|$ as a function of momentum and time for a Kitaev ribbon of size $202\times 700$. The white and black dashed lines show the beginnings and ends of the critical regions respectively. Bright areas stretching over regions in momentum space show eigenvalues close to zero and are indicative of the critical regions. For the quench $\nu:0\to1$ shown in panel (a), additional bright horizontal lines are visible which are the zero modes contributing to $l_B(t)$.} 
\label{fig:evrde}
\end{figure}

\begin{figure}
\includegraphics[width=0.45\columnwidth]{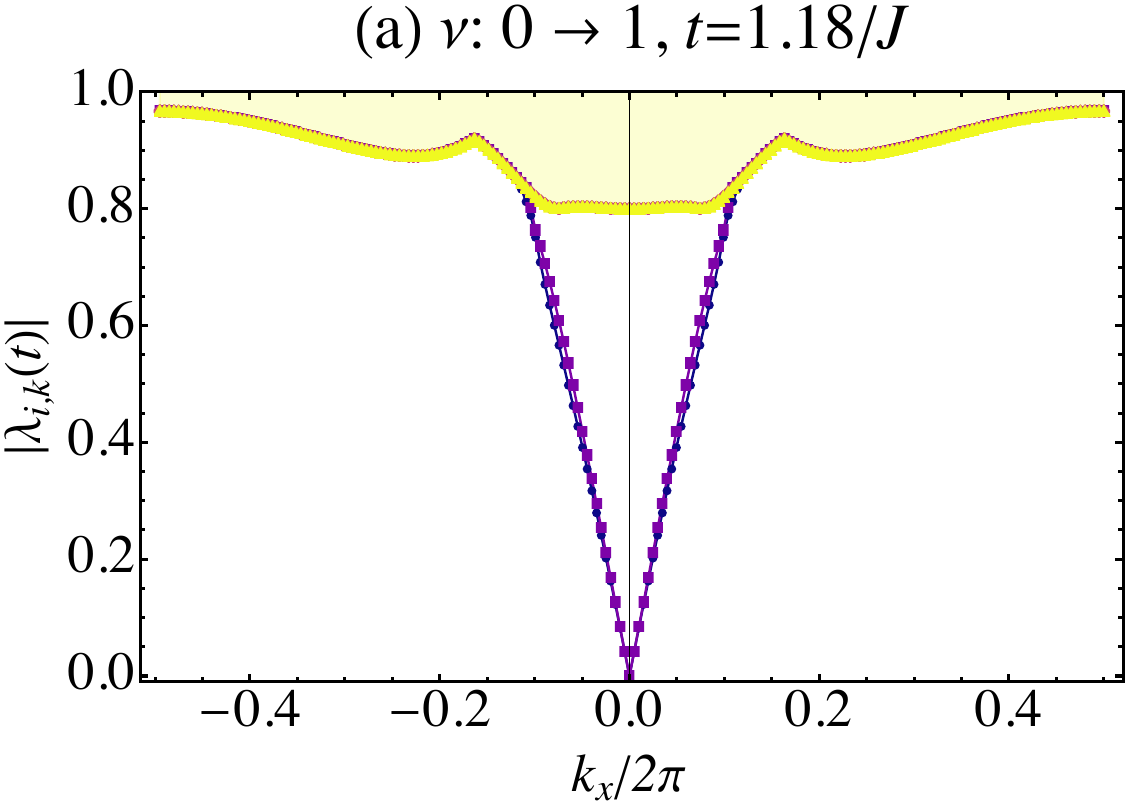}
\includegraphics[width=0.45\columnwidth]{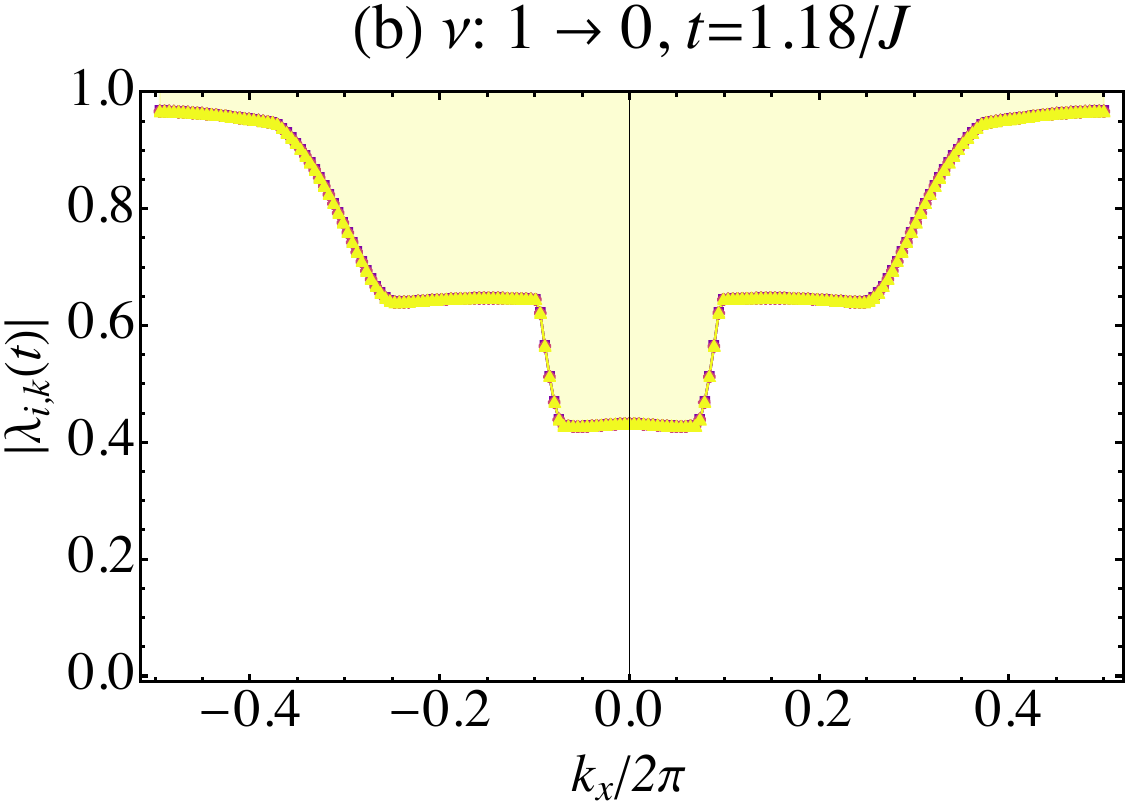}
\caption{The Loschmidt eigenvalues with smallest magnitude at particular times as a function of momentum for the Kitaev ribbon of size $202\times 700$, see also Fig.~\ref{fig:evrde}. Bulk eigenvalues are shown as a light shaded region. The boundary modes are visible as in-gap linear bands in panel (a).} 
\label{fig:evrs}
\end{figure}

Missing from Fig.~\ref{fig:evr} however is any information about the momenta which can help demonstrating that these are indeed in-gap \emph{bands}. In Fig.~\ref{fig:evrde} we therefore plot the lowest eigenvalue $|\lambda_0(t)|$ for the ribbon resolved by its longitudinal momentum $k$. The critical regions are caused by Fisher zeroes crossing the real time axis, which can be clearly seen both for the $\nu:0\to 1$ and the $\nu:1\to 0$ quench. Additionally the bands of in-gap modes which give rise to the boundary term $l_B(t)$ for the $\nu:0\to 1$ quench are now clearly visible in Fig.~\ref{fig:evrde}(a) as horizontal bright lines between gap closing points even after the critical regions have started to overlap. In Fig.~\ref{fig:evrs} we present in addition examples of cuts through Fig.~\ref{fig:evrde}(a,b) at particular times where the in-gap band for the $\nu:0\to 1$ case is also clearly visible. 

As is clear from Fig.~\ref{fig:evrs}, and as we have checked more widely, the in-gap modes have a linear dispersion. It is therefore relatively easy to calculate their contribution to the boundary return rate and thus showing the connection between the Loschmidt spectrum and the boundary contribution to the dynamical free energy in a quantitative manner. Writing the total width of the in-gap dispersing modes as $2\Delta k$, the `velocity' as $v$ and the eigenvalue when they enter the bulk as $\tilde\lambda=v\Delta k$ we have, measuring $k$ from the middle of the in-gap modes, bands like $|\lambda_k|=vk=\tilde\lambda k/\Delta k$. The boundary return rate caused by these modes per edge can then be directly calculated as
\begin{equation}\label{bound_an}
    \tilde l_B=\frac{\delta\Delta k}{\pi}\left(1-\ln\tilde\lambda\right)
\end{equation}
where $\delta$ is the degeneracy of the in-gap modes. If there are exponentially small zero modes $\sim e^{-N_y\alpha_i}$, as in the ribbon case, these contribute a constant term $\alpha_i$ each giving
\begin{equation}\label{bound_an0}
    \tilde l_{B+0}=\frac{\delta\Delta k}{\pi}\left(1-\ln\tilde\lambda\right)+\sum_i\alpha_i.
\end{equation}
We can extract the parameters $\Delta k$, $\delta$, and $\tilde\lambda$ from the numerical data and compare these results to the boundary return rate found from the scaling analysis. The $\alpha_i$ can similarly be found from a scaling analysis. We note that $\Delta k$ gradually decreases between the critical times, which is why there is no plateau for the boundary return rate, but rather a slope downwards (as $\tilde\lambda$ also changes this is not necessarily a simple dependence). The difference between the flake and ribbon boundary return rate should therefore be $\sum_i\alpha_i$. This is confirmed by the symbols shown in Fig.~\ref{fig:br} and more examples can be found in appendix \ref{app:dbb}.

\section{Conclusions}\label{sec:con}

In this article we have provided strong numerical evidence for a dynamical bulk boundary correspondence in two-dimensional topological systems by establishing a direct link between in-gap modes of the non-Hermitian Loschmidt matrix and boundary contributions to the dynamical free energy. The in-gap modes and the connected boundary contributions appear periodically between DQPTs suggesting that they represent an order parameter distinguishing different topological phases. It is this analogy between in gap modes and a putative topology of the non-Hermitian Loschmidt matrix, and the well known equivalent boundary modes and topology of Hamiltonians which prompts the name DBBC.For the accessible time scales we see no sign that this is a transient phenomena. Changes in the clarity of the in-gap modes we ascribe to finite size effects.

While our study is numerical and a full analytical theory of the dynamical bulk-boundary correspondence is so far lacking, these results are important because they do demonstrate that this link, first found in one-dimensional topological systems, is generic and not a consequence of one dimension where eigenvalues of the Loschmidt matrix have no dispersion and are naturally pinned to zero. Indeed, by fitting the dispersion of the in-gap modes we have quantitatively checked this connection and by considering the effects of disorder, see App.~\ref{app:disorder}, we have shown that the in-gap modes are stable thus further supporting their likely topological origin.

Having identified this link numerically, the remaining future task is to define a dynamically changing topological index for the non-Hermitian Loschmidt matrix. Here we note that detailed studies of the topology of two-dimensional non-Hermitian systems have only recently started \cite{Sirker_2dnh}. In addition, it would be desirable to relate the DBBC to other physical quantities which might be easier to observe experimentally than the boundary return rate.

\acknowledgments
N.S~acknowledges support from the National Science Centre (NCN, Poland) under the grant 2024/53/B/ST3/02600. J.S.~acknowledges support by NSERC via the Discovery grants program.  All data used in this article, and some supporting data, can be found at Ref.~\cite{Maslowski2025}.

\appendix

\section{Properties of the Loschmidt matrix}\label{app:prop}

We can make unitary rotations of $\bm M(t)$ in Eq.~\eqref{rle} which will naturally not influence $L_{N_x\times N_y}(t)$ defined in Eq.~\eqref{LN}. For computational purposes it is convenient to choose a basis in which $\hh_0$ is diagonal. Let us assume that only the $f$ lowest states of $\hh_0$ are filled in the initial state and denote the unitary transformation to this basis by $\bm S$. The correlation matrix, $\C$, which acts as the projection operator $\mathbb{P}_f$ onto the filled states of $|\Psi_0\rangle$, then has a simple form. Assuming $\hh_0$ is ordered by ascending energies then
\begin{equation} \label{cin0}
\C = \begin{pmatrix}
    \mathbb{1}_{f} & \mathbb{0} \\
\mathbb{0} & \mathbb{0}_{N-f}
\end{pmatrix},
\end{equation}
where $\mathbb{1}_{f}$ is the $f$--dimensional identity matrix and $\mathbb{0}$ a null matrix. In this basis we find
\begin{equation} \label{min0}
\bm M(t) =\begin{pmatrix}
\E (t) & \B (t) \\
\mathbb{0} & \mathbb{1}_{N-f} \\
\end{pmatrix}
\end{equation}
where $\E(t) = \mathbb P_f\, \bm S\, \exp(-it {\mathcal H}^1) \bm S^\dagger\, \mathbb P_f $ and $\B(t) = \mathbb P_f\, \bm S\, \exp(-it {\mathcal H}^1) \bm S^\dagger (\mathbb{1}- \mathbb P_f) $. To find the eigenvalues of $\bm M(t)$ therefore requires only the calculation of $\bm S$ and $\exp(-it {\mathcal H}^1)$. $\bm M(t)$ has $N-f$ eigenvalues equal to one and we find
\begin{equation} \label{la}
L(t)= \det \bm M(t) = \det \E(t) \,.
\end{equation}
If the bands are half filled, as in the cases considered in this article, then $f=N/2$.

\section{Bandstructures}\label{app:bands}

\begin{figure}
\includegraphics[width=0.45\columnwidth]{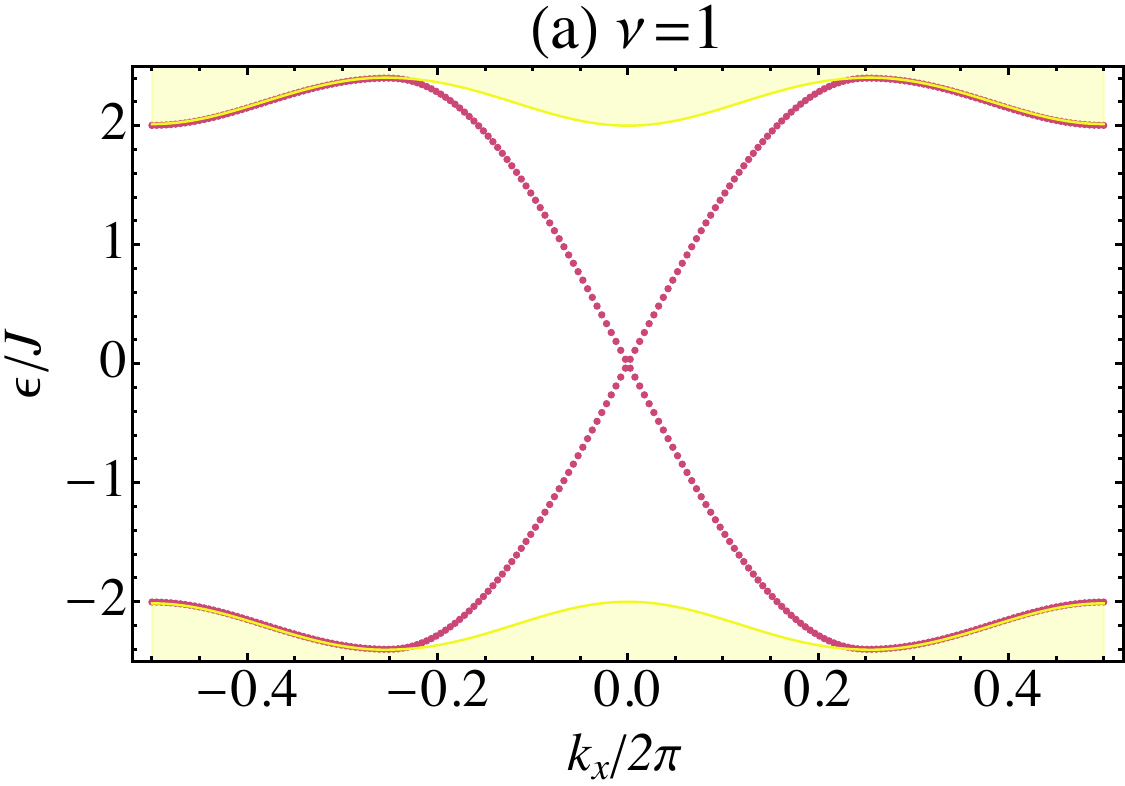}
\includegraphics[width=0.45\columnwidth]{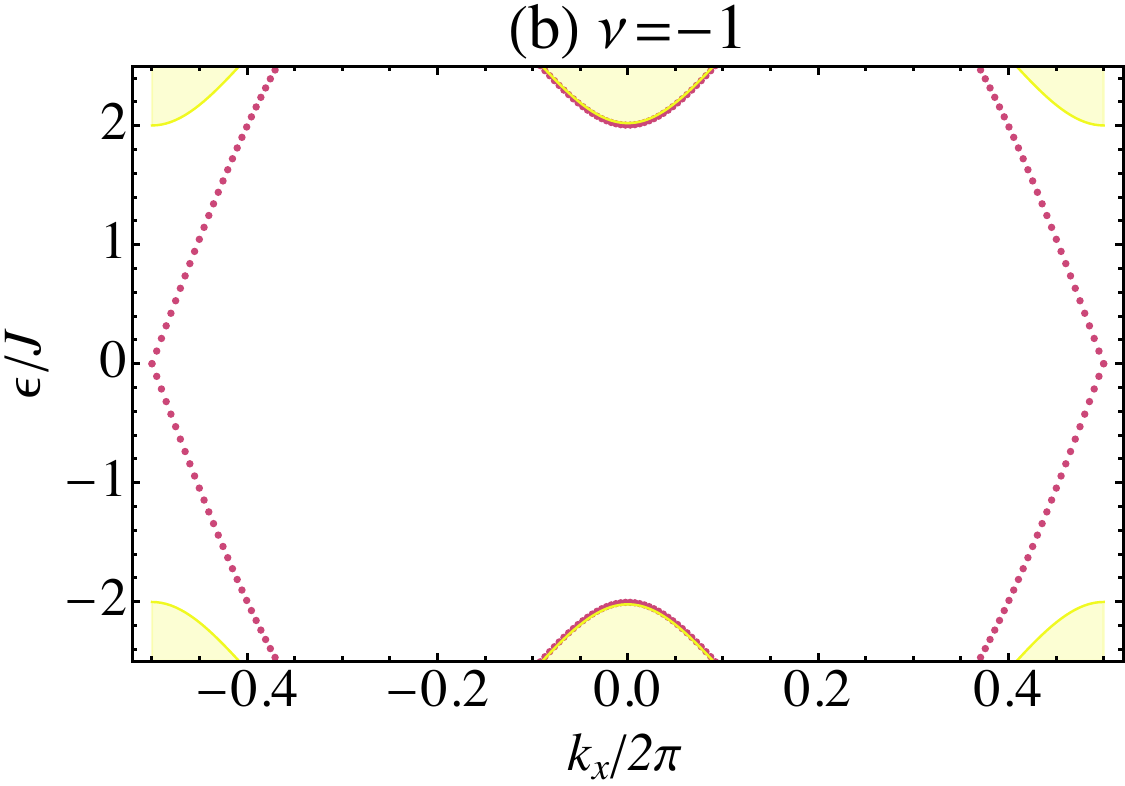}
\caption{Bandstructures for the Kitaev ribbon at two points indicated in Fig.~\ref{fig:phase} with (a) $\nu=1, \Delta>0$, and (b) $\nu=-1$. The topologically protected bands can be seen crossing the gap. Left and right moving bands are on opposite edges of the ribbon.} 
\label{fig:bs}
\end{figure}
Fig.~\ref{fig:bs} shows the bandstructures for the two topologically non-trivial phases of the Kitaev model. The Hamiltonian is given in Eq.~\eqref{kit_ham}.  Each has a chiral band with the band crossing at either $k=0$ or $k=\pi$. Note that the chirality of the edge modes is opposite for the two phases.

\section{Scaling examples}\label{app:scaling}

\begin{figure}
\includegraphics[width=0.9\columnwidth]{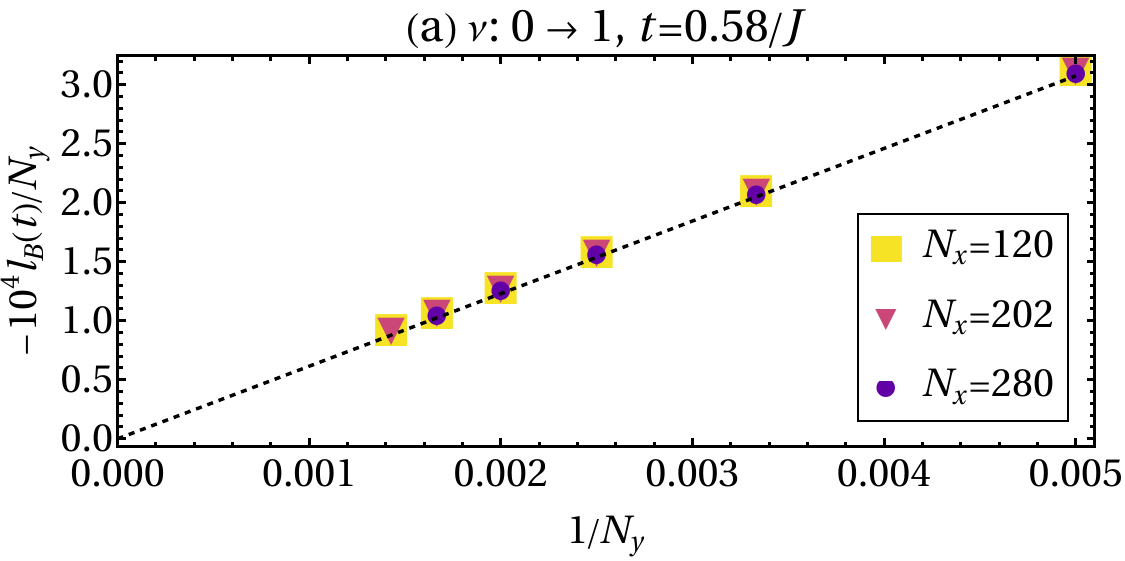}\\
\includegraphics[width=0.9\columnwidth]{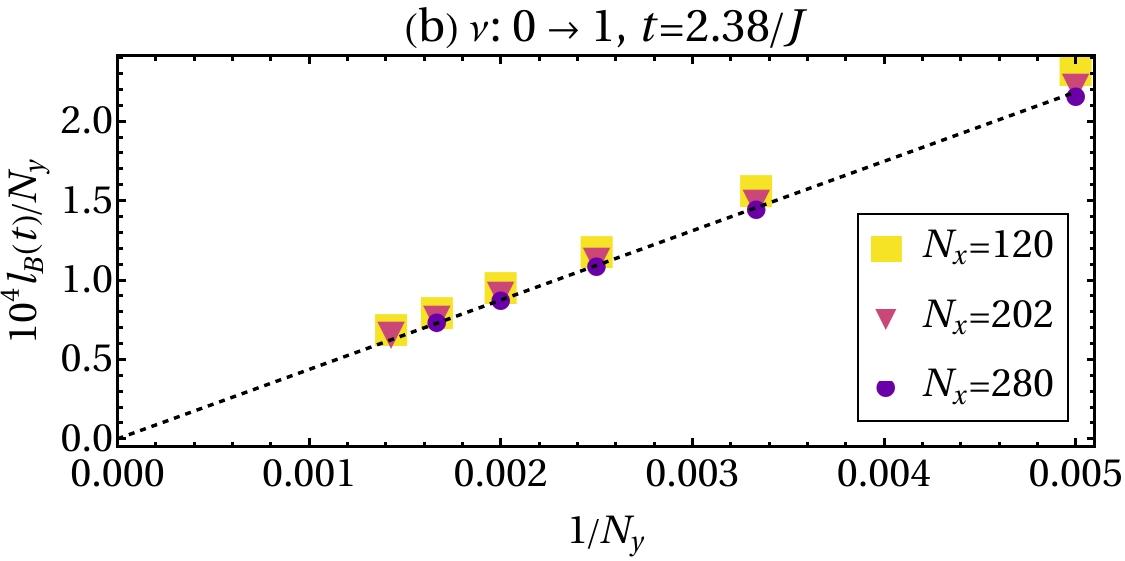}
\caption{Here we show examples of the scaling for the ribbons. We take two distinct times for one of the quenches, as marked on the plots. The dashed line shows the fitted boundary contribution $l_B(t)$ as extracted from the $N_x=202$ data according to Eq.~\eqref{bbreturn}. $A(t)$ is extracted form the scaling for the flakes as given by Eq.~\eqref{bbreturno}. Discrete points are given by Eq.~\eqref{scalecheck} with $N_x$ and $N_y$ as marked.} 
\label{fig:kitsc}
\end{figure}

To demonstrate the validity of the scaling ansatz defined in Eq.~\eqref{bbreturn} and Eq.~\eqref{bbreturno} which leads to the results shown in Fig.~\ref{fig:kitsc} we give examples of the boundary contribution extracted by the scaling for $N_x=202$ and compare to
\begin{equation}\label{scalecheck}
    l_{N_x\times N_y}(t)-l(t)-\frac{A(t)}{N_xN_y}.
\end{equation}
In the limit that Eq.~\eqref{bbreturn} holds this should be approximately $l_B(t)$. As can be seen, this is well observed, and we also show several data points for different $N_x$ and $N_y$ to further demonstrate the validity of Eq.~\eqref{bbreturn}.

\section{Influence of disorder}\label{app:disorder}
\begin{figure}
\includegraphics[width=0.9\columnwidth]{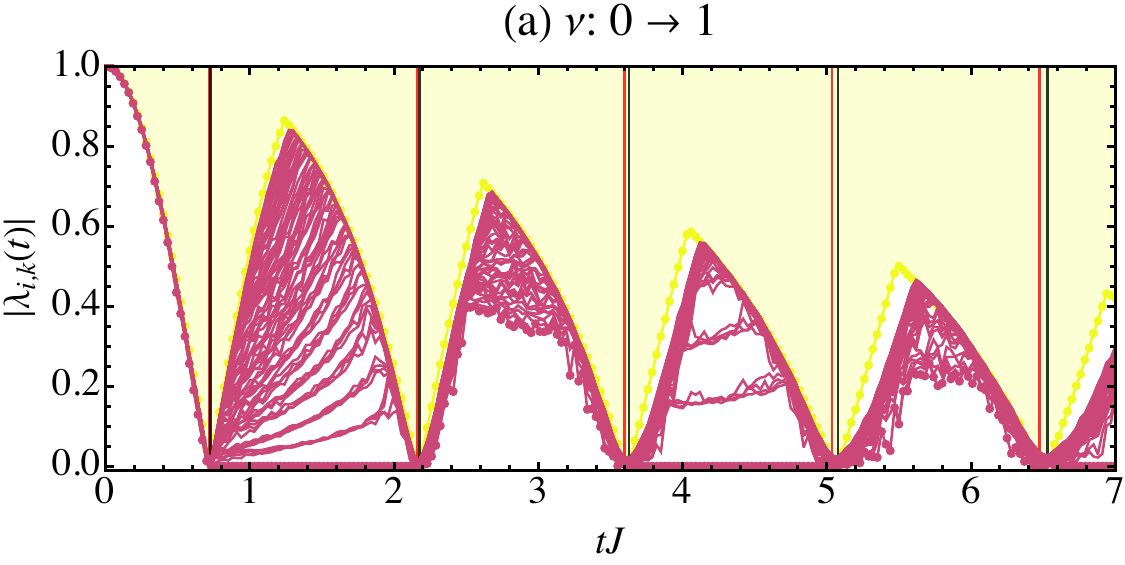}\\
\includegraphics[width=0.9\columnwidth]{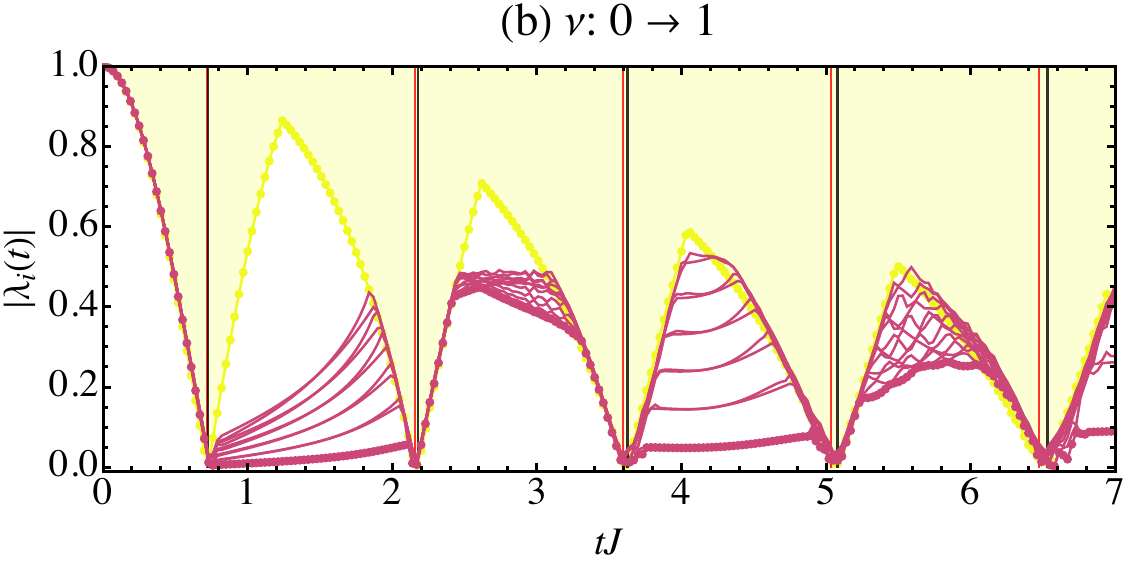}
\caption{Same as Fig.~\ref{fig:evr} but with potential disorder of up to 10\% added to the chemical potential. Shown is a single disorder realization in each case. While small variations are visible, the in-gap states are robust.} 
\label{fig:kitdis}
\end{figure}

Currently, a full analytical theory of the DBBC is lacking. In order to give some further evidence that the phenomenon is indeed topological in origin we here check the robustness of the in-gap modes to the influence of disorder. For the ribbon we add a random term to the chemical potential which varies along $x$ and is constant along $y$, and fluctuates up to 10\% of the value of the actual chemical potential. For the flake the random term fluctuates onsite everywhere. In Fig.~\ref{fig:kitdis}(a,b) we show one disorder realization for each geometry. We checked multiple disorder realizations for both the ribbon and flake, and also considered disorder in the on-site pairing in addition to the case of disorder in the chemical potential. In all cases we have found that the in-gap states are robust supporting the interpretation that they are of topological origin. Data for further examples can be found at Ref.~\cite{Maslowski2025}.

\section{Further examples}\label{app:dbb}

\begin{figure}
\includegraphics[width=0.95\columnwidth]{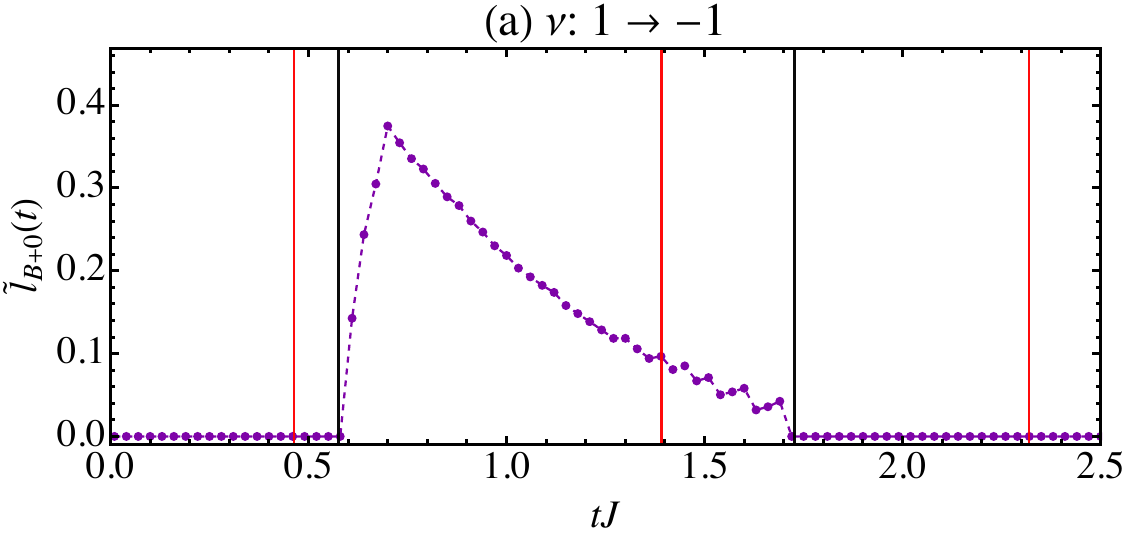}\\
\includegraphics[width=0.95\columnwidth]{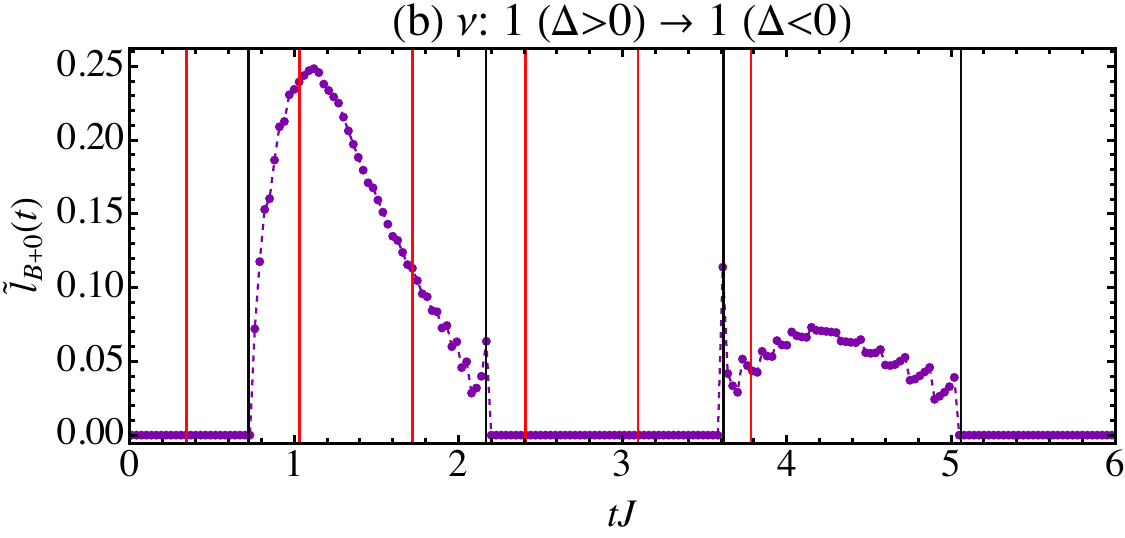}
\caption{Boundary return rate $\tilde l_{B+0}(t)$ for a Kitaev ribbon calculated by fitting the in-gap bands, see Eq.~\eqref{bound_an0}. A direct comparison to $l_B(t)$ is hindered by insufficient system sizes to perform a stable scaling analysis.} 
\label{fig:br2}
\end{figure}

\begin{figure*}
\includegraphics[width=0.95\columnwidth]{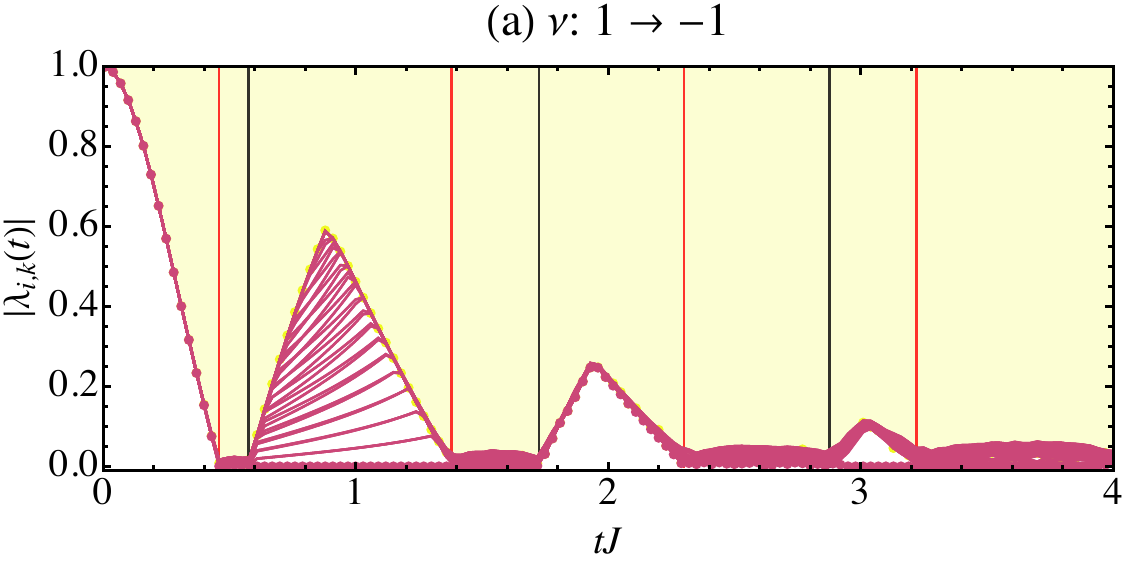}
\includegraphics[width=0.95\columnwidth]{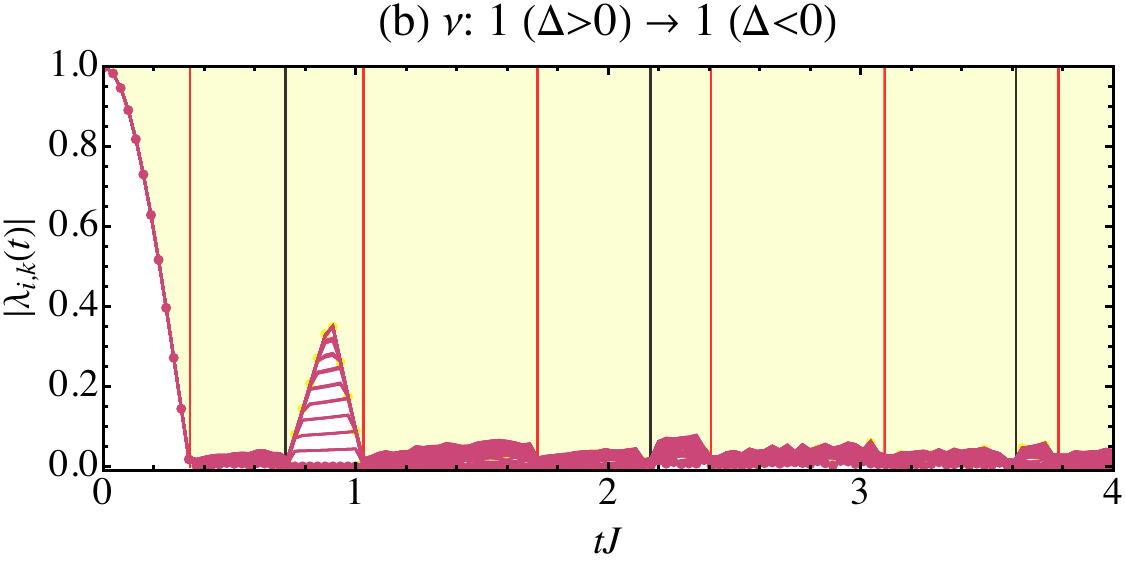}
\caption{(a,b) 80 smallest Loschmidt eigenvalues $|\lambda_i(t)|$ for a Kitaev ribbon of size $202\times 700$. The shaded regions show the areas where there are bulk eigenvalues. Vertical red and black lines show the beginnings and ends of critical regions respectively. In-gap eigenvalues are visible for short times as expected based on the DBBC, but critical times soon overlap making it difficult to see whether or not these modes also exist at later times. This issue is resolved in Fig.~\ref{fig:evd2} and Fig.~\ref{fig:evs2}, see below.} 
\label{fig:ev2}
\end{figure*}

Here we show examples of two more quenches which have DQPTs: $\nu\!:1\to-1$ and $\nu\!:1\to1$ but switching the sign of $\Delta$. This second case does not cross an equilibrium phase boundary, but does still have DQPTs. In Fig.~\ref{fig:br2} we show the boundary return rate for these two quenches, as found from an analysis of the in-gap eigenvalues, see Eq.~\eqref{bound_an0}. For these two quenches the critical regions soon overlap, which makes it hard to distinguish the in-gap modes at longer times. In both cases a boundary term can be clearly seen at short times, even for the quench within the same non-trivial phase. Fig.~\ref{fig:ev2} shows the Loschmidt spectrum for the ribbon geometry, which shows the expected boundary modes. Whether they are  periodically appearing and disappearing is numerically difficult to resolve.

\begin{figure}
\includegraphics[height=0.325\columnwidth]{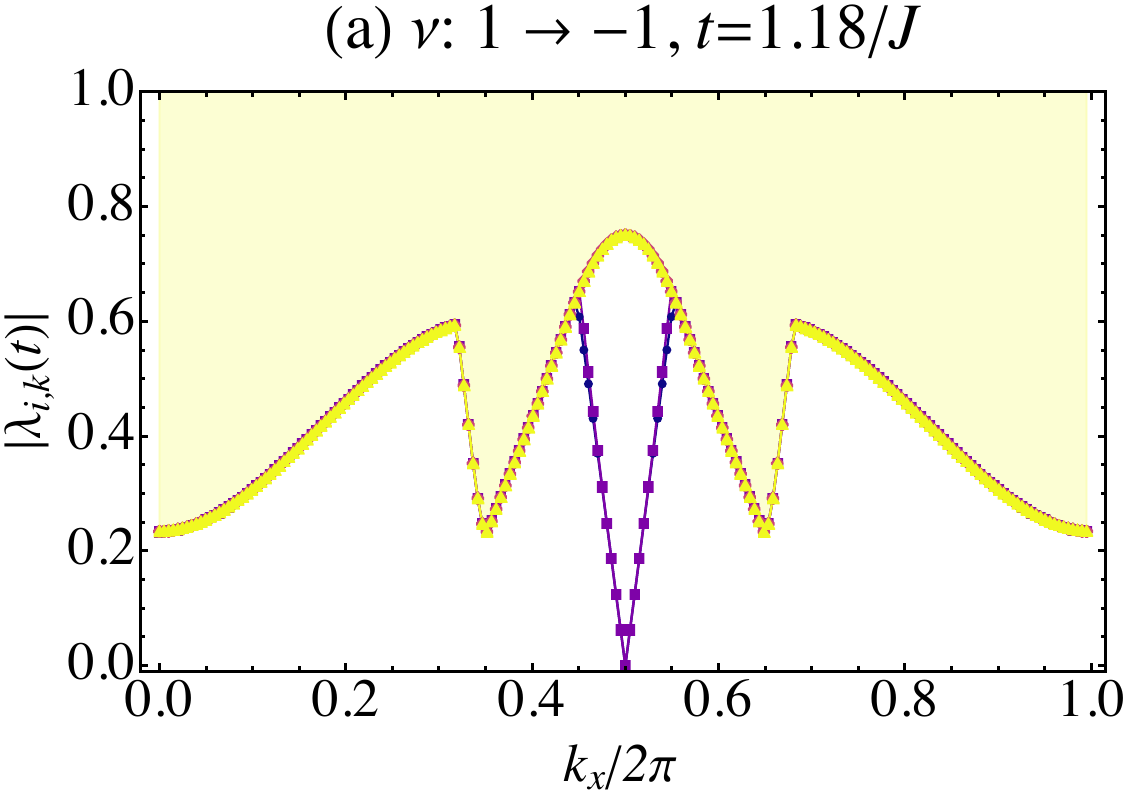}
\includegraphics[height=0.325\columnwidth]{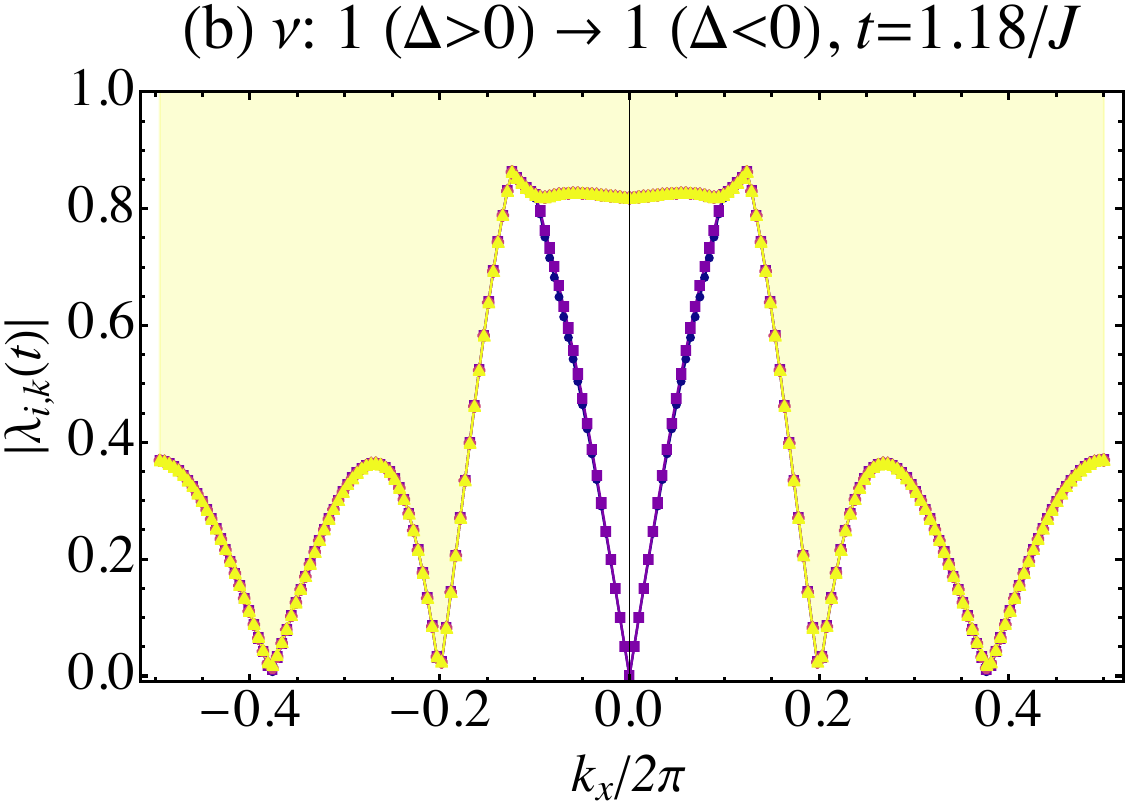}
\caption{Loschmidt eigenvalues of smallest magnitude at particular times as a function of momentum for the Kitaev ribbon of size $202\times 700$. Bulk eigenvalues are shown as a light shaded region. When the shaded region touches zero the system is inside a critical region. The boundary modes are visible as in-gap linear bands.} 
\label{fig:evs2}
\end{figure}

In Fig.~\ref{fig:evd2} we show $|\lambda_{0,k}(t)|$ for these quenches for both the bulk and the ribbon case. A comparison reveals the additional boundary modes in the ribbon case. The shapes of $t_c(\vec{k}^*)$, which give rise to the critical regions and hence to the critical times, can also be clearly seen. For longer times these shapes become stretched, leading to the overlapping of the critical regions.
\begin{figure}
\includegraphics[height=0.405\columnwidth]{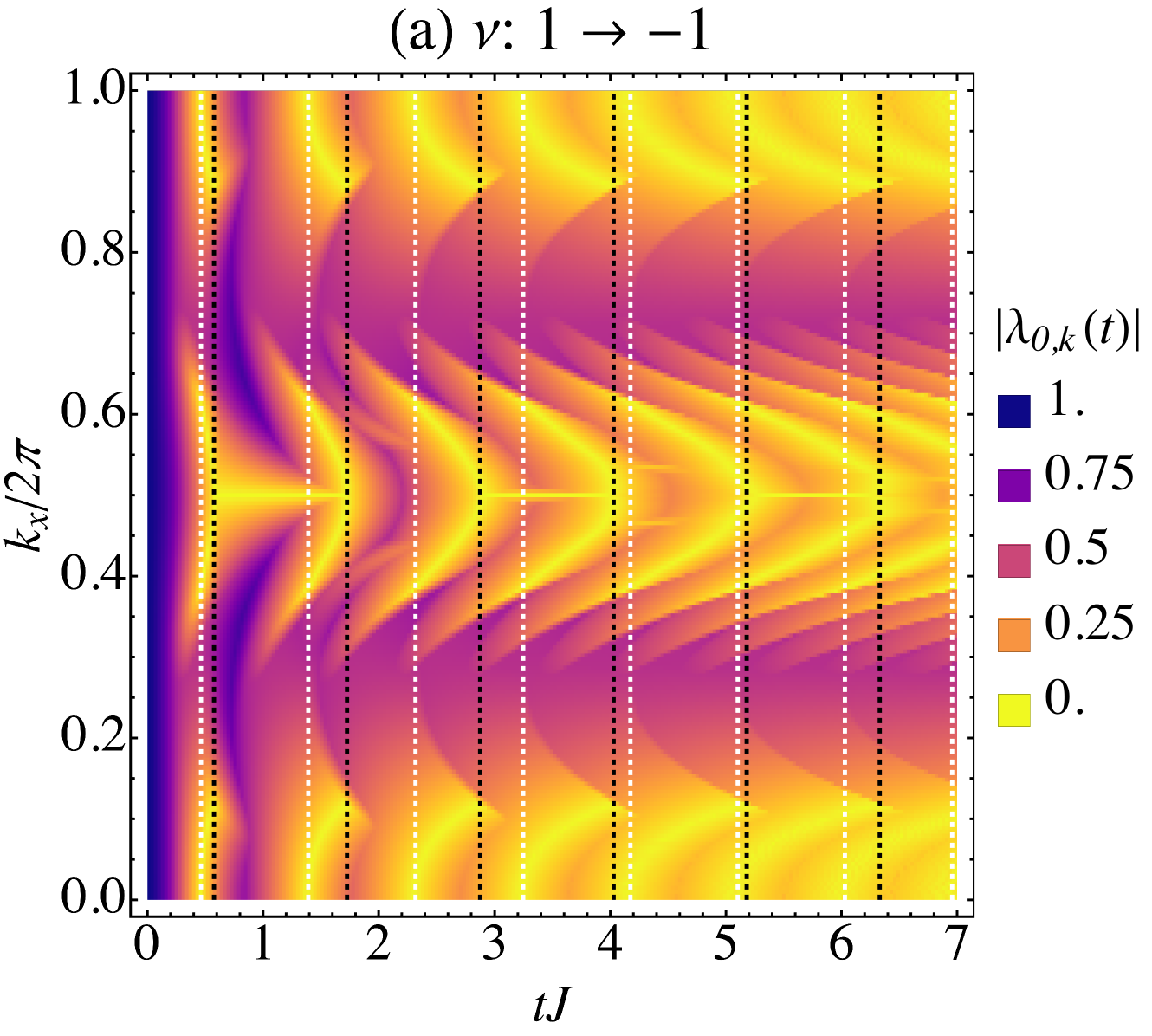}
\includegraphics[height=0.405\columnwidth]{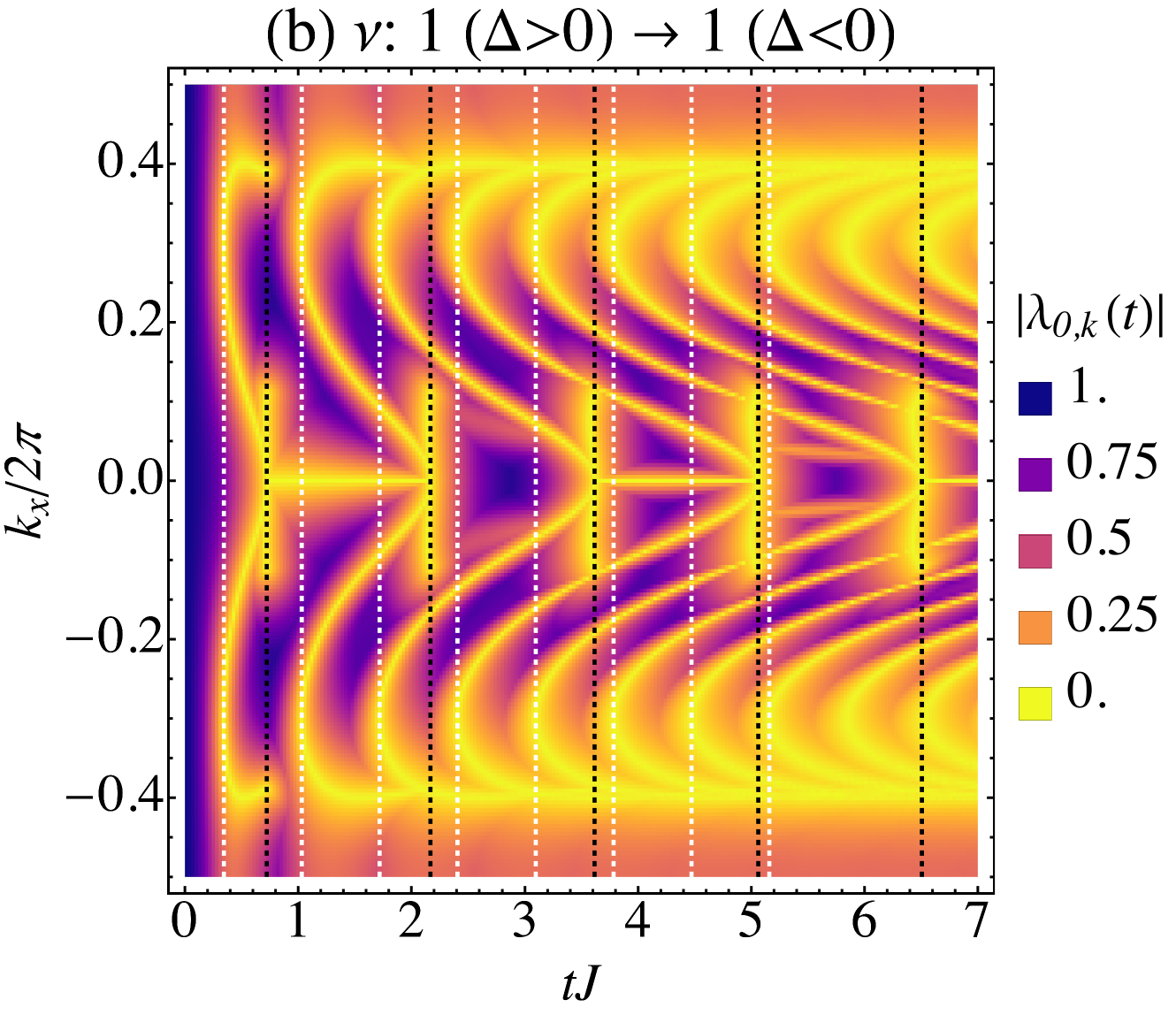}\\
\includegraphics[height=0.405\columnwidth]{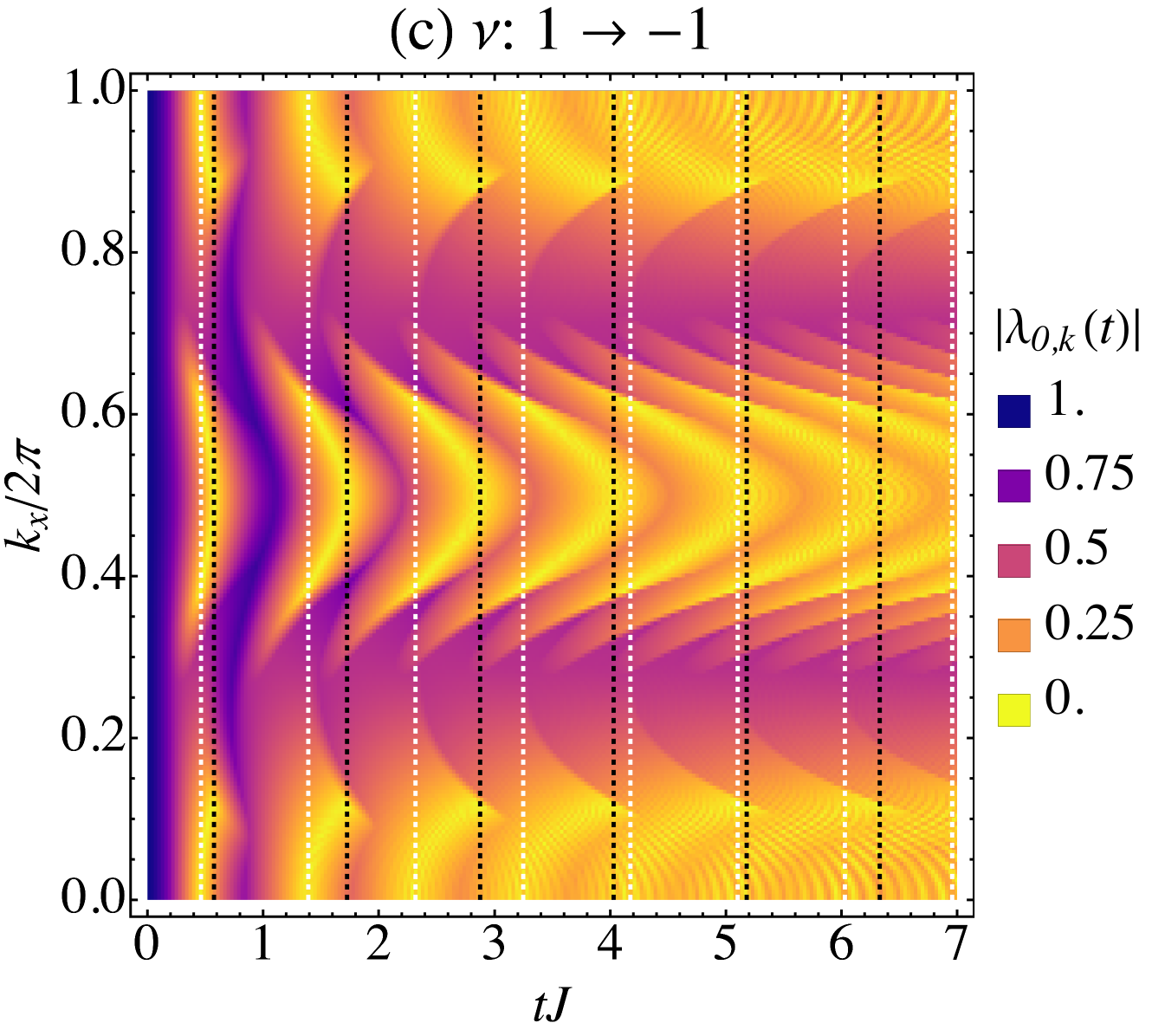}
\includegraphics[height=0.405\columnwidth]{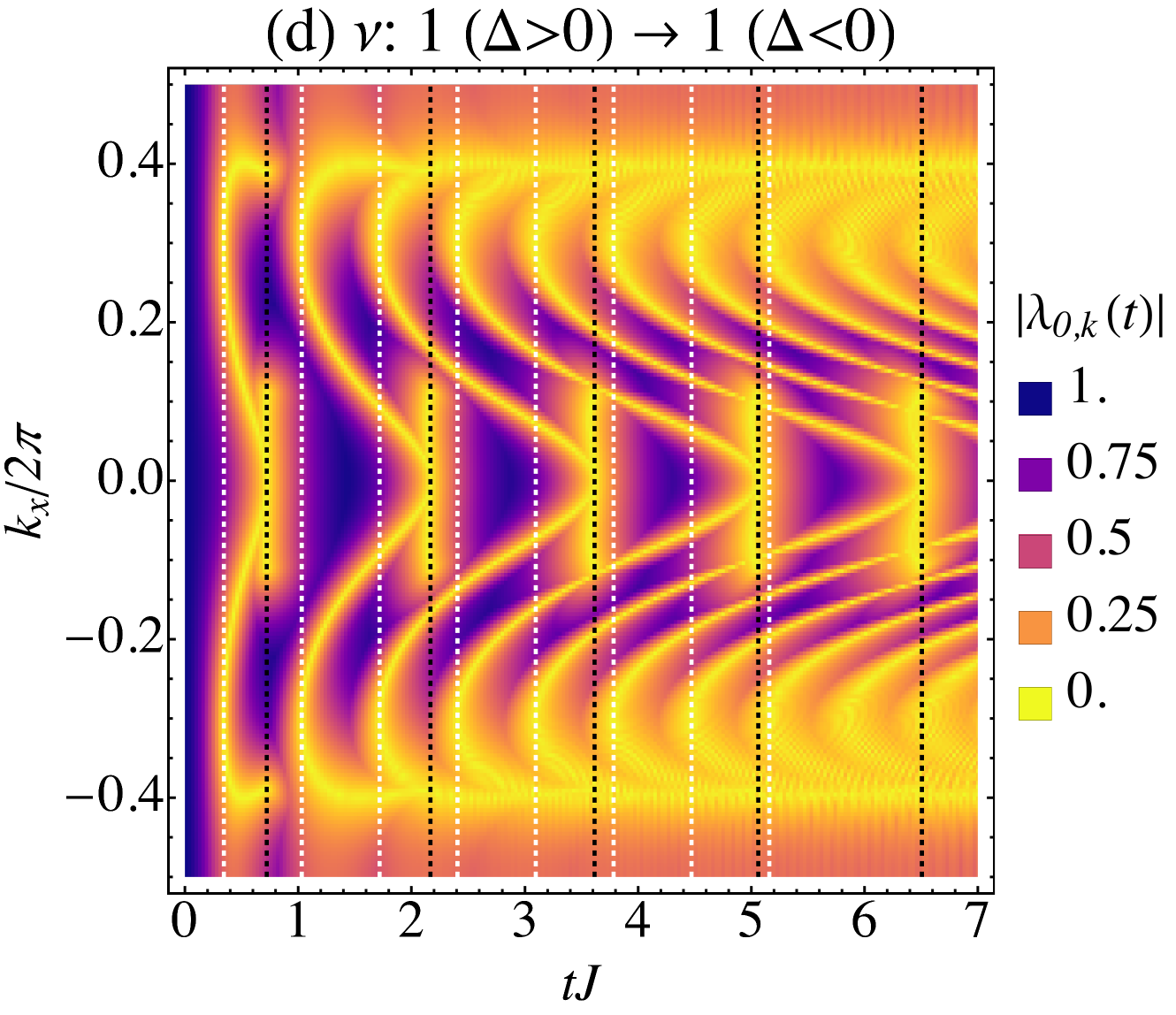}
\caption{The smallest Loschmidt eigenvalue $|\lambda_{0,k}(t)|$ as a function of momentum and time for the Kitaev ribbon of size $202\times 700$, see panels (a,b). In panels (c,d) bulk results are shown for comparison, making it possible to identify the boundary contributions in (a,b).} 
\label{fig:evd2}
\end{figure}
Fig.~\ref{fig:evs2} shows momentum resolved cuts at particular times of Fig.~\ref{fig:evd2}. That the in-gap eigenvalues form bands ``crossing'' the gap is again clearly visible.


%

\end{document}